\documentclass[twoside,12pt]{article}

\usepackage{epsfig}

\usepackage{graphicx}
\usepackage{amsmath}
\usepackage{amssymb}
\usepackage{array}
\usepackage{wrapfig}
\usepackage{units}
\usepackage{caption}
\usepackage{slashed}

\def\bm{\boldmath}

\newcommand{\be}{\begin{equation}}
\newcommand{\ee}{\end{equation}}
\newcommand{\bea}{\begin{eqnarray}}
\newcommand{\eea}{\end{eqnarray}}

\newcommand{\ba}{\begin{eqnarray}}
\newcommand{\ea}{\end{eqnarray}}

\newcommand{\nc}{\newcommand}

\nc{\newsection}[1]{\section{#1}\setcounter{equation}{0}}
\nc{\newappendix}[1]{\section*{#1}\setcounter{equation}{0}}
\nc{\scm}{\scriptscriptstyle\mathrm}
\nc{\f}{\frac}
\nc{\baa}{\begin{array}}      \nc{\eaa}{\end{array}}
\nc{\bit}{\begin{itemize}}    \nc{\eit}{\end{itemize}}
\nc{\ben}{\begin{enumerate}}  \nc{\een}{\end{enumerate}}
\nc{\bce}{\begin{center}}     \nc{\ece}{\end{center}}
\nc{\bfl}{\begin{flushright}} \nc{\efl}{\end{flushright}}
\nc{\btb}{\begin{tabular}}    \nc{\etb}{\end{tabular}}
\nc{\eps}{\varepsilon}
\nc{\vp}{\varphi}
\nc{\tvp}{\widetilde{\varphi}}
\nc{\D}{\mbox{$\not\!\!D$}}
\nc{\Db}{\mbox{${\raisebox{2mm}{\boldmath ${}^\leftarrow$}\hspace{-4mm} D}$}}
\nc{\Dfb}{\mbox{$\raisebox{2mm}{\boldmath ${}^\leftrightarrow$}\hspace{-4mm} D$}}
\nc{\vpj }{\mbox{${\vp^\dag i\,\raisebox{2mm}{\boldmath ${}^\leftrightarrow$}\hspace{-4mm} D_\mu\,\vp}$}}
\nc{\vpjt}{\mbox{${\vp^\dag i\,\raisebox{2mm}{\boldmath ${}^\leftrightarrow$}\hspace{-4mm} D_\mu^{\,a}\,\vp}$}}
\newcommand\slurp[1]{#1}
\newcommand\sslurp[2]{#1{#2}}
{\catcode`/=\active \expandafter}%
\slurp{\newcommand/}{/\penalty1000\hskip0pt\relax}
{\catcode`:=\active \expandafter}%
\slurp{\newcommand:}{:\penalty1000\hskip0pt\relax}

\newcommand\addspace{\ifcat\nextchar a\spacefactor999. \else.\fi}
{\catcode`\.=\active \expandafter}%
\slurp{\newcommand.}{\futurelet\nextchar\addspace}

\usepackage[unicode]{hyperref} 
\ifx\href\undefined\def\href#1{}\fi
\ifx\texorpdfstring\undefined\def\texorpdfstring#1#2{#1}\fi

\newcommand\myslash{/} \newcommand\mycolon{:}
\newcommand\doi{{\catcode`/=\active \catcode`:=\active \expandafter}\sslurp\realdoi}
{\catcode`/=\active \catcode`:=\active \expandafter}%
\slurp{\newcommand\realdoi[1]{{\let/=\myslash \let:=\mycolon
                               \edef\raw{{http://dx.doi.org/#1}}\expandafter}%
                               \expandafter\href\raw{doi:#1}}}
\newcommand\eprint[2]{{\escapechar-1%
                       \edef\a{\expandafter\string\csname arXiv\endcsname}%
                       \edef\b{\expandafter\string\csname #1\endcsname}%
                       \edef\c{\expandafter\string\csname #2\endcsname}%
                       \edef\d{\noexpand\href{http://arXiv.org/abs/\c}}%
                       \ifx\a\b\expandafter\d\fi{\tt #1:#2}}}

\begin{document}

\title{ \vspace{1cm} 
Dark Matter Studies Entrain Nuclear Physics 
}
\author{
Susan Gardner$^1$ and 
George Fuller$^{2}$
\\
\\ 
$^1$  Department of Physics and Astronomy, University of Kentucky, \\  Lexington, KY 40506-0055, USA
\\
$^2$ Department of Physics, University of California, San Diego, \\ La Jolla, CA 92093, USA 
}

\maketitle
\begin{abstract} 
We review theoretically well-motivated dark-matter 
candidates, and pathways to their discovery, in the 
light of recent results from collider physics, astrophysics, and cosmology. 
Taken in aggregate, these encourage broader thinking 
in regards to possible dark-matter candidates --- dark-matter need not
be made of ``WIMPs,'' {\it i.e.}, elementary particles with weak-scale masses and interactions. 
Facilities dedicated 
to nuclear physics are well-poised to investigate certain 
non-WIMP models. 
In parallel to this, developments in observational cosmology 
permit probes of the relativistic energy density 
at early epochs and thus provide 
new ways to constrain dark-matter models, provided nuclear physics inputs
are sufficiently well-known. 
The emerging confluence of accelerator, astrophysical, and cosmological constraints 
permit searches for dark-matter candidates 
in a greater range of masses 
and interaction strengths than heretofore possible. 
\end{abstract}

\newpage

\section{Introduction} 

A key problem in modern physics is the nature of the dark matter, and many facets of this issue overlap significantly with current theoretical and experimental efforts in nuclear physics. 
There is no doubt that much of the mass-energy 
content of the universe is dark and resides in as yet unknown forms. 
Disjoint astronomical observations provide 
compelling evidence for the existence of 
additional, non-luminous matter, or dark matter, in 
gravitational interactions. 
The current evidence includes the pattern of acoustic oscillations in the power spectrum of the
cosmic microwave background (CMB)~\cite{Hu:2001bc}, 
the relative strength and shape of the galaxy-distribution 
power spectrum at large wave numbers~\cite{Eisenstein:2005su}, as well as 
observations of long-standing of galactic rotation curves at 
distances for which little luminous matter is present~\cite{Faber:1979pp,Rubin:1980zd}. 
The cosmological evidence, taken collectively, implies that dark matter comprises 
some twenty-three percent of the energy density of the universe today,
with a precision of a couple of percent~\cite{Hinshaw:2012}.
Independent threads of observational evidence show that 
we live in a dark-dominated universe, with studies of 
Type Ia supernovae revealing the existence
of dark energy~\cite{Riess:1998cb,Perlmutter:1998np}. 
The dark sector is 
diverse in that it separates into distinct dark matter and dark 
energy components, and the individual components themselves 
may also be of diverse origin. 
The complexity of the known universe gives such a possibility appeal, 
though Occam's razor argues for the simplicity of a single dark-matter component. 
Indeed, it has been long thought that dark matter could be explained by an as yet
undiscovered, massive, weakly interacting elementary particle, a ``WIMP,'' 
though we cannot currently say 
whether dark matter is comprised of particles of any sort. 
Alternatives to the usual 
cosmological paradigm of dark energy and dark matter 
appear to have less observational 
support~\cite{Clowe:2006eq}, though observational tests 
continue~\cite{Bean:2010zq,Lombriser:2010mp,Weinberg:2012es} and are 
well-motivated as long 
as particle dark matter remains undiscovered.

A Weakly Interacting Massive Particle (WIMP) is still a leading
candidate to comprise the bulk of the dark matter. 
In part this is because there is a robust 
prediction of this particle's contribution to the relic 
energy density 
based on the mass of the particle, its 
weak interaction cross section, and its 
attendant well-determined temperature scale at 
which it would fall out of equilibrium in the early universe. 
The compatibility of this estimate with the observed energy density
in dark matter  is the WIMP ``miracle.'' Most known particles, 
{\it e.g.}, baryons and leptons, do not obey this relation 
between cross section, mass, and relic density. 
Nevertheless, we believe the criterion is 
more properly regarded as a simple 
test of whether a particular particle can be a 
credible candidate for 
a significant component of the dark matter. 
WIMPs have other appealing aspects. 
The lightest supersymmetric particle
might well be the WIMP and, as a 
consequence, WIMPs can have a natural connection to the physics
being probed in current collider experiments. 
Arguably though, the most attractive attribute of WIMPs 
is that they can be directly detectable. 
Their expected densities and spatial distributions 
in the Galaxy combined with their weak interactions
position them for detection via several clever technologies. 
Direct detection searches have not produced a completely compelling signal,
though the next generation of detector technologies is 
poised to push into WIMP mass and cross-section regimes 
where these particles may yet be found. 

Much effort has been invested in devising and testing 
particle-physics models of dark matter, particularly those connected 
to the physics of the electroweak scale. The efforts in this direction
have been recently and thoroughly 
reviewed, noting, {\it e.g.},
Refs.~\cite{Bertone:2004pz,Gaitskell:2004gd,Jaeckel:2010ni,Feng:2010gw,Feng:2010tg}. 
Nevertheless, very recent experimental results and observational measurements 
suggest important shifts in perspective which this review imbues. 
The recent discovery of a 
Higgs-like particle of 125 GeV 
in mass~\cite{atlas:2012gk,cms:2012gu} has 
immediate implications for models of the weak scale and 
for would-be dark-matter models as well. 
Simple models with technicolor are ruled out, 
and the Higgs mass appears to be uncomfortably heavy 
for some popular models with 
minimal weak-scale supersymmetry, note, {\it e.g.}, Ref.~\cite{Feng:2012rn}. 
The limits on the superpartner masses in such models also 
continue to strengthen, though they can be evaded~\cite{Murayama:2012jh}. 
Dark-matter candidates 
in supersymmetry continue to be well-motivated, but 
the recent experimental results prompt thinking of a broader compass, 
and models without weak-scale masses and interactions can successfully
confront the observed relic density~\cite{Feng:2008ya}. 
Paralleling these developments are exciting new opportunities 
for the study of particle physics 
from observational cosmology. The advent of 
precision cosmology and, particularly, the determination of 
a precise value of the baryon-to-photon 
ratio $\eta$ from studies of the cosmic microwave 
background (CMB), 
with the promise of a sub-1\% precision determination from 
Planck~\cite{Planck:2006aa,Ade:2011ah}, 
promotes the study of the light element 
abundances from big-bang nucleosynthesis (BBN) 
to an exquisite probe of 
physics beyond the standard model (BSM) of particle physics and 
of non-standard cosmology~\cite{Steigman:2007xt}. 
At issue is the possibility of ``late'' energy injection in the evolving
early universe, most plausibly from the decay 
of weakly-coupled matter, possibly of 
dark matter or its familiars. 
Thus strongly-coupled probes of 
new physics at the LHC act in counterpoint 
to observational probes of the weakly-coupled cosmos, 
and an era in which we have powerful, 
complementary probes of new physics is upon us. 

Neutrinos are a known part of the weakly interacting universe, and their 
interactions as well as intrinsic nature are also probed by BBN constraints
--- note Ref.~\cite{Steigman:2012ve} for a recent review. 
Yet the BBN constraints realized from 
confronting the observed primordial element abundances, mindful of possible 
contamination from nonprimordial sources, with theoretical predictions 
are only one of several possibilities. 
We can also probe the energy density associated with weakly interacting sources
directly by measuring the expansion rate of the early universe. 
In standard big-bang cosmology, 
the expansion rate is controlled by the Friedmann equation, namely, by the time-evolution
of the Hubble constant $H(t) \equiv {\dot a}/{a}$,
where $a(t)$ is the scale factor.
We define the instantaneous closure parameter to be 
$\Omega(t)\equiv 8\pi G\rho(t)/3H^2$, with 
Newton's gravitational constant $G$ and the energy density $\rho(t)$. 
Contributions to $\Omega(t)$ can be
codified by their scaling behavior in $a(t)$, so that as $t\to -\infty$, 
the contribution
with the highest inverse power in 
$a(t)$ dominates --- consequently, that from relativistic
species, or ``radiation,'' dominates the energy budget at the earliest times. Photons, neutrinos,
as well as relativistic electrons and positrons can contribute to it. 
Studies of the CMB at the epoch of photon decoupling 
can also limit the sum of the neutrino masses $\Sigma_i m_{\nu_i}$~\cite{Abazajian:2011dt}, and 
the comparison of this result to terrestrial 
studies
could reveal new physics, {\it e.g.},  
the existence of sterile neutrinos~\cite{Fuller:2011qy}. 
Observations of the small and large scale 
structure of the cosmos are also key probes of dark matter. 
The various constraints act both to limit non-standard-model neutrino interactions as 
well as to probe various models of dark matter. The ability to separate the possibilities
is under ongoing development; 
it is possible that new physics in the interactions of 
known neutrinos could be confused with evidence for dark matter~\cite{Harnik:2012ni}.
Nevertheless, hints and signals as to the nature of 
dark matter can be inferred not only from the 
interplay of terrestrial and cosmological neutrino mass limits~\cite{Fuller:2011qy}, 
but also from the observed departure from the expected 
relativistic energy density at the CMB epoch,  
as well as from a failure to confront the predictions of BBN. 

Terrestrial studies of 
neutrons and nuclei play a key role in the interpretation of these 
cosmological tests, making the emerging picture 
of the cosmos from these studies an additional concrete 
outcome of such measurements. 
For example, the theoretically predicted light-element abundances 
from big-bang cosmology rely on measured nuclear reaction cross sections
and the neutron lifetime. 
Despite the maturity of the subject, discussion and measurement of these fundamental 
quantities continue, in part because the terrestrial cross section measurements
have not always been made at the center-of-mass energies relevant to BBN 
conditions~\cite{Boyd:2010kj}. 
The ${^4{\rm He}}/{\rm H}$ abundance 
is ultimately set by the neutron-to-proton ratio in the BBN 
epoch. 
This is controlled by the neutron lifetime in the standard model. 
Interestingly, 
the ${^4{\rm He}}$ yield is particularly sensitive 
to a possible lepton asymmetry, specifically 
the electron neutrino and electron antineutrino imbalance, as well as to the 
relativistic particle energy density.
The recent foment over the proper value 
of the neutron lifetime~\cite{greene} has yielded 
a shift in its assessment by the PDG from 
$\tau_n=885.7 \pm 0.8 \,{\rm s}$
to 
$\tau_n=880.1 \pm 1.1 \,{\rm s}$~\cite{PDG2012}. 
This yields a small but appreciable reduction 
in the ${^4{\rm He}}/{\rm H}$ 
abundance of ${\cal O}(0.001)$, and the resolution of this shift, if not 
yet observationally practical, is important in principle because behind it could lurk a nonzero
lepton asymmetry, namely, in $\nu_e$ and $\bar\nu_e$ --- as well as information on 
the relativistic particle energy density. 
No method yet exists to probe the lepton asymmetry terrestrially, though the observation 
of neutrinoless double $\beta$-decay would 
change its interpretation; it would reflect an imbalance in neutrino chirality, rather than
a particle-antiparticle asymmetry. 
Nollett and Holder~\cite{Nollett:2011aa} point out that improved 
measurements of the ${{\rm D}}/{\rm H}$ abundance could also yield insight on BSM physics
and cosmology, but the ${^4{\rm He}}/{\rm H}$ abundance 
is intrinsically more sensitive to
relativistic particle energy density and a lepton  asymmetry~\cite{Steigman:2012ve}. 
Although the D/H yield in BBN is much less sensitive 
to energy density and new BSM neutrino physics than is that of 
$^4{\rm He}$/H, if the primordial 
deuterium abundance can be measured accurately 
enough it could provide insights into, 
and competitive constraints on, BSM issues~\cite{Smith:2006uw}.

Ongoing cosmological observations can give
us fresh insights about dark matter, though we should emphasize all that we know now 
about its properties, 
as well as its existence, comes from astrophysics and cosmology. 
Observations of large-scale structure tell us that 
dark matter must be stable, or at least metastable,  on Gyr time scales. 
Moreover, dark matter cannot be 
``hot'' at the redshift at which it 
decouples from matter in the cooling early Universe~\cite{White:1984yj}. 
Here we use temperature, {\it i.e.}, whether
it is ``cold'' or ``hot'' to connote whether its thermal energy makes its 
non-relativistic or relativistic, respectively, at decoupling. 
For so-called thermal relics, this criterion selects 
the mass of the dark candidate as well,
so that colder particles are heavier. However, alternative production 
scenarios can exist, and very light particles
can also act as cold dark matter, as in the case of the axion~\cite{Sikivie:2006ni}. 
Finally, dark matter appears to be weakly interacting, so that it 
appears to lack both electric and color 
charge, 
though infinitesimally charged dark matter is not completely excluded. 
The evidence in broad brush speaks to a universe with cold, collisionless dark matter; 
this in concert with dark energy as a cosmological constant
gives rise to the $\Lambda$CDM paradigm.\footnote{We note ``CDM'' is 
cold dark matter, see Ref.~\cite{Blumenthal:1984bp}.}
 We will, however, be more 
broad-minded in our
description and consider warm, weakly self-interacting dark-matter models as well.

We begin our review in earnest with a more 
detailed description of what has been established
observationally thus far in regards to dark matter, 
as well as an extended prospectus of what yet may come. 
We then survey 
a spectrum of DM models, which we regard as well-motivated because 
they happen to resolve more problems than simply giving 
identity to a dark-matter candidate. 
Enormous effort has been devoted to the study of dark matter and to the construction of
models which can describe it. A comprehensive review of this vast literature is beyond
the scope of our planned article; rather we select such topics which connect to 
facilities and expertise which exist in nuclear physics. 
We consider {\it supersymmetric} models, whose 
motivation lie in their connection to the resolution of
the hierarchy problem. Such models have been thoroughly reviewed~\cite{Feng:2010gw},  
so that we are more concerned with offering an overview of the broader 
possibilities, supplemented with a discussion of the computation and impact of certain 
 needed hadron matrix elements.
We pay particular attention to {\it hidden sector} models, in which dark matter dynamics
are controlled by an internal gauge symmetry. In such models, the stability of 
dark matter
is explained if it carries a hidden conserved charge. 
The hidden gauge bosons can
potentially be probed through precision fixed-target experiments 
at intermediate energy facilities for nuclear physics, such as at 
JLab and MAMI, 
or through refined measurements of the $g-2$ of the muon. 
We also consider 
{\it asymmetric} models, whose motivation lie in their explanation of 
why the dark-matter and matter relic densities are commensurate in size. 
We round out our review with a discussion of 
sterile neutrino models of dark matter, 
which connect 
naturally to a relativistic energy density at the photon decoupling epoch in 
excess of standard-model predictions. 
We believe that were 
the existence of light, sterile neutrinos established in terrestrial experiments, 
a role for sterile neutrinos in the resolution of the dark-matter problem would become
more strongly motivated. 
We note in passing that axion 
models are a very well-motivated 
class of models which resolve the strong CP problem, but we eschew detailed
discussion of them here, noting that excellent reviews 
of that topic already exist~\cite{Jaeckel:2010ni,Feng:2010gw}. 
As appropriate we include limits on dark-matter
models from dark-matter direct and indirect detection efforts, noting that 
the physics reach of single 
experiments depend on particular astrophysical inputs, as
well as assumptions in regards to dark-matter--matter interactions.

\section{Dark Matter from Observations}

Observational studies of the large-scale structure of the Universe, in concert
with numerical simulations, as well as 
studies of galaxies and galactic clusters, constrain the nature of dark matter. We summarize
these emergent, gross features because viable 
particle-physics models of dark 
matter must be compatible with them. 
In particular, in the context of standard Big-Bang cosmology, whether
dark matter is hot or cold, that is, whether it is 
relativistic or not 
in the epoch at which it is sufficiently cool to decouple from its interactions
with ordinary matter, 
impacts the formation of large-scale structure after the Big Bang. 
In the scenario in which dark matter is formed as a thermal relic in the cooling early Universe, 
this criterion also selects the mass of the candidate particle. 
If dark matter is cold and collisionless, then 
galaxy formation proceeds via a hierarchical clustering~\cite{Press:1973iz,White:1977jf}, 
namely, from the merging of small protogalactic clumps on ever larger
scales; and this is supported by numerical simulations~\cite{Blumenthal:1984bp,Davis:1985rj}.
In contrast, if dark matter is hot, the hierarchy is inverted, so that 
large protogalactic disks, or ``pancakes''~\cite{zeldovichp},
form first and then break into clumps~\cite{Bond:1980ha,White:1984yj,Bond:1988ez}. 
Galaxies, however, are observed at much larger redshifts than the latter simulations
predict~\cite{White:1984yj,Bond:1988ez}. Moreover, 
observations of particular classes of quasar absorption 
lines, the so-called damped Lyman-$\alpha$ systems,
thought to be the evolutionary progenitors of galaxies
today, also favor a cold-dark-matter scenario~\cite{Kauffmann:1995gi,Prochaska:1997xi}. 
It has also been argued that hot dark matter, {\it i.e.}, most notably, 
light, massive neutrinos, 
cannot explain the galactic rotation curves~\cite{Tremaine:1979we}. 
However, the cold-dark-matter 
paradigm also generates significant clumpiness below the Mpc scale, so that 
a galaxy the size of the Milky Way should host many satellite subhaloes and 
indeed many observable satellite 
galaxies --- many more than observed~\cite{Kauffmann:1993gv,Klypin:1999uc,Moore:1999nt}. 
Recent discoveries of very faint Milky
Way dwarf galaxies suggest that the problem could be, at least in part, of an 
observational origin; we refer to Ref.~\cite{Bullock:2010uy} for a review and further discussion. 
Warm dark matter has also been advocated as a way to 
alleviate these difficulties~\cite{Bode:2000gq,Moore:1999gc,AvilaReese:2000hg}. 
Limits on the mass of 
warm dark matter emerge from the comparison of the observations 
of the Lyman-$\alpha$ absorption spectrum with numerical 
simulations~\cite{Viel:2005qj,Seljak:2006qw,Viel:2006kd,Viel:2007mv,Boyarsky:2008xj}; 
the limits depend on the particle considered 
and the manner in which it is produced~\cite{Petraki:2007gq}, 
yielding, {\it e.g.}, a candidate mass $M > 12.1$ keV for a nonresonantly produced, 
thermal energy spectrum 
sterile neutrino at Bayesian 95\% confidence interval~\cite{Boyarsky:2008xj}. 

Additional cosmological constraints exist on the mass of a 
dark-matter particle, in the event that it is produced as a {\it thermal relic}. 
If the particles annihilate via the 
weak interaction, then $\sigma_{ann} v$  
is parametrically set by ${\cal N}_AG_F^2 M^2$, 
where $G_F$ is the Fermi constant, 
${\cal N}_A$  is a dimensionless factor, 
and we assume 
$\sigma_{ann} \propto 1/v$.
In this case avoiding a dark-matter abundance in excess of 
the observed relic density bounds $M$ from below.  
Indeed, 
under these conditions the mass of the cold dark-matter particle must exceed 
${\cal O}(2\,\hbox{GeV})$ to avoid closing the 
Universe~\cite{Hut:1977zn,Lee:1977ua,Vysotsky:1977pe}. 
The resulting  lower bound  on $M$ can be relaxed in different ways. 
Feng and Kumar~\cite{Feng:2008ya}, {\it e.g.}, have emphasized that 
the appearance of $G_F$ in $\sigma_{ann} v$ is simply parametric, 
that $G_F$ can be replaced with $g_{\rm eff}$, and that 
the effective coupling $g_{\rm eff}$ can be small without
having the precise numerical value of $G_F$. Thus if $g_{\rm eff} > G_F$, 
the bound on $M$ is weakened. 
Indeed, such considerations permit dark matter candidates which 
confront the relic density and big-bang nucleosynthesis 
constraints successfully but 
range from the keV to the TeV scale in mass~\cite{Feng:2008ya,Feng:2008mu}. 

We know other things about dark matter. For example, 
the consistency of the determinations of the fraction of the energy density of the universe 
in dark matter today suggest that dark matter must be at least  metastable over roughly 10 Gyr
time scales, though 
the anomalies noted 
at PAMELA~\cite{Adriani:2008zr,Adriani:2010ib} and Fermi~\cite{FermiLAT:2011ab} 
in the positron fraction of
cosmic rays for energies in excess of roughly 20 GeV 
probe the possibility of decaying dark matter 
today~\cite{Arvanitaki:2009yb}. 
Moreover, dark-matter self-interactions have been suggested as a way of alleviating 
some problems with the cold-dark-matter hypothesis at galactic distance 
scales~\cite{Spergel:1999mh}. However, we also 
know that dark matter cannot have an appreciable strong~\cite{Dimopoulos:1989hk,Gould:1989gw}
or electromagnetic~\cite{Gradwohl:1992ue} charge, so that we can, in zeroth
approximation, regard dark matter as collisionless. 

The broad features which emerge from this summary are 
that dark matter is either cold or warm, stable or metastable, and 
lacks substantial self-interactions, via a strong or electromagnetic charge. 
Viable dark-matter models must be compatible with these features. 
However, these constraints do not 
preclude \lq\lq secret,\rq\rq\ 
non-standard model self-interactions 
among dark matter particles, 
and these have been 
suggested as a way to explain observed Milky Way satellite galaxy 
morphology~\cite{Rocha:2012jg,Peter:2012jh}.

\section{A Prospectus of Cosmological Constraints}

There are 
exciting new possibilities for the 
experimental and observational study of light, weakly coupled degrees of freedom,  
setting up a tightly constrained, 
nearly over-determined situation where physics beyond the standard model (BSM)
may well show itself. 
Such studies may ultimately point to BSM neutrino interactions or to a modification 
of standard, big-bang cosmology, but they can also have implications for the nature of the dark sector. 

Dark matter and dark energy together comprise some 95 percent
of the closure or critical 
density.
It is possible to determine its fractional components. 
For example, we know the baryon density 
from the observations of the ratio of the amplitudes of the 
acoustic peaks in the cosmic microwave background (CMB) 
radiation.
This measurement corroborates the Big Bang 
Nucleosynthesis (BBN)-based determination 
of the baryon density from the deuterium 
abundance as measured in isotope-shifted 
hydrogen absorption lines in high 
redshift gas clouds along lines of sight to 
Quasi-Stellar Objects (QSOs)~\cite{Kirkman:2003uv}. 
The baryon rest mass contribution 
to closure is modest, with a fit derived from the WMAP9 CMB data 
in one case \cite{Hinshaw:2012} 
yielding $\Omega_b = 0.0463 \pm 0.0024$. In short, the baryon 
rest mass contributes about 20 percent of the 
non-relativistic dark matter content of the universe today. 
Neutrinos have small rest masses, and 
they may contribute a smaller fraction of closure as we will describe. 

The total kinetic plus gravitational potential energy of the contents 
inside an arbitrary two-sphere, co-moving with the 
expansion in the universe, appears to be very close to zero. 
Expressed in terms of the fractional contribution to the 
closure energy density today, 
$\Omega_k$, this energy is very small, consistent with zero. 
We note, {\it e.g.}, 
that $\Omega_k=-0.001 \pm 0.012$ from CMB data alone~\cite{Hinshaw:2012}.

In summary, the universe appears to be flat, {\it i.e.}, with curvature parameter 
$k=0$, to fair precision. This is significant because $k=0$ is a fixed value in the evolution of the universe, a spacetime symmetry. This condition corresponds to a total mass-energy density 
always equal to the instantaneous critical value. 
Total $\Omega$, once set to unity, must always be unity, {\it regardless of how the microphysics might transform mass-energy in the universe from one form into another}. Put another way, once established, $\Omega=1$ will persist so long as the microphysics operating in the universe respects a key symmetry condition: at any time the overall spatial distribution of mass-energy must be homogeneous and isotropic. 

The significance of this symmetry for the dark sector is at once obvious and profound: there is nothing in gravitation or spacetime physics itself to argue against there being many kinds of particles and other entities carrying mass-energy that contribute to $\Omega=1$. We already know that there are several components to the dark sector, as we have 
described.
One way that the dark sector is
diverse is that it separates into distinct dark matter and dark 
energy components. Given that the spacetime symmetry implied by $\Omega=1$ is blind to how the mass-energy is divided up among components,
there is nothing to preclude the individual components themselves 
from being of diverse origin, with many kinds of dark matter and even dark energy. 
This perspective makes it particularly natural to consider a role for neutrinos
in the dark sector, too. 

Terrestrial experiments have told us 
the neutrino mass-squared differences and three ($\theta_{1 2}$, $\theta_{2 3}$, $\theta_{1 3}$) 
of the parameters in the unitary transformation between neutrino energy (mass) 
states and the weak interaction (flavor) states~\cite{PDG2012}. 
Setting aside $CP$-violating phase(s), we lack only knowledge of 
the neutrino mass hierarchy, though it is of particular import for astrophysics, 
both for core collapse supernovae, 
and for the \lq\lq measurement\rq\rq\ of the neutrino mass through cosmological 
observations~\cite{Abazajian:2011dt}. 
Future and planned observations
promise sensitivity to the sum of neutrino masses 
at the $0.1\,{\rm eV}$ scale and smaller ~\cite{Abazajian:2011dt};  in this regard 
the prospect of the detection of the weak gravitational lensing of the CMB shows particular promise. 
Consequently, since the sum of the light neutrino masses should exceed $0.05\,{\rm eV}$ 
in the normal mass hierarchy and $ 0.1\,{\rm eV}$ in the inverted mass hierarchy such observations
should be able to resolve the neutrino mass 
hierarchy and, in essence, provide a detection of the relic neutrino background. 
This would be a remarkable discovery, not least for its new window on the dark sector. 
We know this relic background must be present at the epoch of weak freeze-out in Big Bang Nucleosynthesis (BBN), 
$T\sim 1\,{\rm MeV}$, else we would not get the agreement that we have between BBN predictions 
and the observationally-inferred primordial abundances of deuterium and ${^4{\rm He}}$. 
Nevertheless, the relic density and/or energy spectra of the neutrinos 
between the BBN epoch 
and the decoupling of photons at $T_{\rm CMB} \approx 0.2\,{\rm eV}$ can be modified by new physics, particularly
by particle decay, and observations, not just of $\sum m_\nu \ne 0$,  can
limit this possiblity. 

The imprint of a generation of particles which have decayed away can be inferred not only from the 
interplay of terrestrial and cosmological neutrino mass limits~\cite{Fuller:2011qy}, 
but also from the observed departure from 
the expected relativistic energy density at the CMB epoch, 
as well as from a failure to confront the predictions of BBN. 
For the moment we consider the latter two mechanisms explicitly. 
The next generation CMB experiments and telescopes will be able to provide 
relatively precise bounds on the energy density of particles 
with relativistic kinematics at the epoch ($T_{\rm CMB}$) of 
photon decoupling. By convention, this \lq\lq radiation\rq\rq\ energy density is parameterized as follows:
\begin{equation}
\rho_{\rm rad} = {\left[  2+\frac{7}{4} {\left( {\frac{4}{11}} \right)}^{4/3}\, N_{\rm eff}  \right]}\,
{\frac{\pi^2}{30}}\,T_{\rm CMB}^4 .
\label{Neff}
\end{equation}
Standard model 
physics robustly predicts $N_{\rm eff} \approx 3.046(1)$~\cite{Mangano:2005cc}. 
The excess over $3$, 
corresponding to three flavors of neutrinos 
with black body, Fermi-Dirac-shaped energy spectra, arises from 
$e^\pm$-pair annihilation into out-of-equilibrium 
neutrino pairs near and during the BBN epoch. 
It is important to note that $N_{\rm eff}$ parameterizes {\it all} relativistic energy density 
at the photon decoupling epoch, not just that contributed specifically by the known active neutrinos. 
Any measurement of $N_{\rm eff}$ significantly different from $3.046$, either lower or higher, 
signals new physics, either new particle physics, or some deviation in the 
history of the early universe from that predicted by the 
standard model.

Current CMB measurements of $N_{\rm eff}$ are not very precise, 
but consistent with the 
standard model; 
they are, nevertheless, 
tantalizing to some. 
For example, the South Pole Telescope 
reports $N_{\rm eff} = 3.71 \pm 0.35$ (quoting $1\,\sigma$ errors)~\cite{Hou:2012xq}, 
employing 
both Hubble parameter and Baryon Acoustic Oscillation priors, while 
WMAP9 reports  $N_{\rm eff} = 3.84 \pm 0.40$~\cite{Hinshaw:2012}, 
and the Atacama Cosmology Telescope collaboration reports 
$N_{\rm eff} = 2.78 \pm 0.55$~\cite{Sievers:2013wk}. 
All of these measurements are 
consistent with the standard model value within $2\sigma$. The Planck satellite and 
future CMB polarization observations, by contrast, should give $N_{\rm eff}$ 
to better than $10\%$ precision~\cite{Ade:2011ah}. This will greatly 
heighten the prospects that this measurement will be able to constrain or signal new physics.
For example, the 
neutrino reactor anomaly and the Mini-BooNE experiment can be interpreted as implying the existence of a light (mass $\sim 1\,{\rm eV}$) sterile neutrino or neutrinos with significant vacuum mixing with active neutrino species. Were this interpretation 
to be correct, it would imply ramifications for $N_{\rm eff}$, in that it would be closer
to $4$ than to $3$, and BBN.
And therein lies another way in which new physics is being boxed-in by observations. BBN predictions of light element abundances {\it also} depend on relativistic energy density, and specifically the energy spectra of $\nu_e$ and $\bar\nu_e$, all in ways different than, but complementary to the way $N_{\rm eff}$ depends on these quantities. The CMB acoustic peak amplitude ratios have given us a rather precise value of the baryon-to-photon ratio, $\eta \approx 6.11\times{10}^{-10}$, and the Planck mission promises to get this number to $\pm 0.74\%$ or better. This, coupled with the increasingly precise determinations of the primordial deuterium abundance, show us that the basic nuclear and weak interaction physics of BBN are well understood and, in broad brush, operate closely along the lines of what standard cosmology predicts. There are some problems, the 
 ${^7{\rm Li}}$ and ${^6{\rm Li}}$ yields, for example; 
and these discrepancies have been argued to be signals of new physics, specifically signaling post-BBN cascade nucleosynthesis stemming from, {\it e.g.}, super-WIMP decay. However, there may be more prosaic explanations of these issues, and the real clincher may be the primordial helium abundance.

Though linear regression with compact blue galaxies yields a primordial helium abundance with very small statistical errors, some believe that there could be significant systematic errors in this approach. Thus, 
right now, for example, the linear regression-inferred helium abundance on its own is not widely viewed as ruling out a light sterile neutrino. However, the next generation of CMB experiments will be able to infer the primordial ${^4{\rm He}}$ abundance from the Silk damping tail on the CMB power spectrum. In essence, 
the more baryons that are locked up in alpha particles as neutrons, 
the fewer electrons there will be, and the longer 
will be 
the 
photon mean free path at the CMB decoupling epoch --- it is this quantity to which 
the CMB measurements are sensitive. CMB-Pol may be able to measure the primordial helium abundance to better than $2\%$. The situation is not perfect because, for example, the primordial helium abundance and $N_{\rm eff}$ are somewhat degenerate. Nevertheless, the prospects for precise helium and $N_{\rm eff}$ constraints are tantalizing. 

Should there be evidence for light sterile neutrinos which stands up against, 
or shows itself in these new cosmological observations, the prospects that 
sterile neutrinos play a role in dark matter will be increased in the 
eyes of many. Likewise, decaying massive particles invoked to satisfy current collider constraints, or 
invoked for lithium production and tied to WIMP dark matter, may possibly leave telltale 
evidence that could be ferreted out with these observations. 

\section{Dark Matter Models}

The standard model 
leaves many questions unanswered: it explains, {\it e.g.}, neither why the weak scale has
the value it has, 
nor the baroque pattern of fermion masses and mixings seen in Nature,
nor the size of the observed baryon asymmetry of the Universe (BAU). 
Most notably, in our current context, it fails to explain dark matter.

\begin{figure}[t!]
\centering
\includegraphics[width=0.6\textwidth]{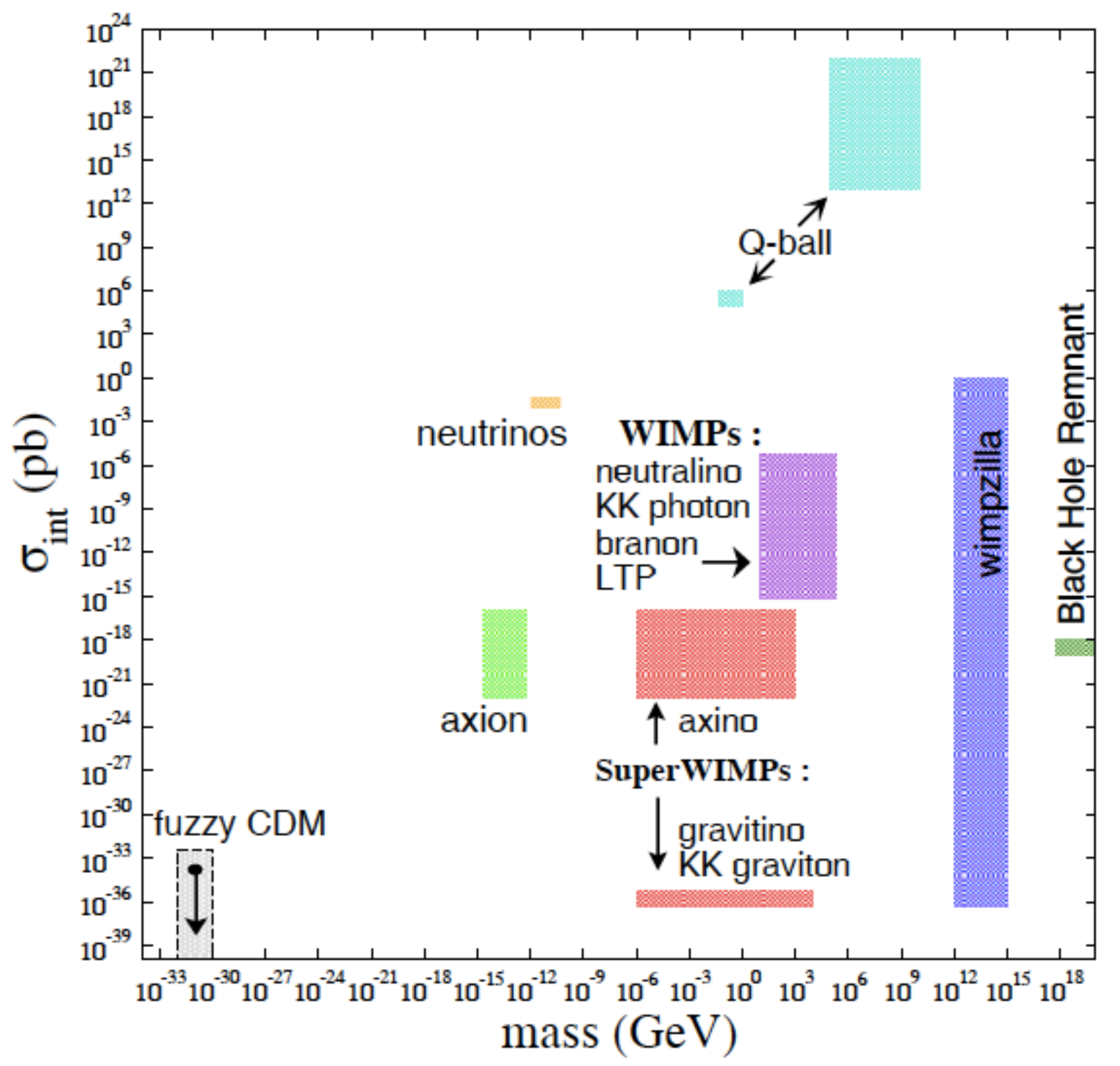}
\begin{minipage}[t]{16.5 cm}
\caption{
Estimated loci of select dark-matter models in the space of candidate mass in GeV versus 
dark-matter-candidate--nucleon interaction cross section in pb, 
figure taken from Ref.~\cite{dmsag2007}. 
}
\label{fig:DMcand}
\end{minipage}
\end{figure}

At the same time, the interpretation of astrophysical observations tells us 
that the needed dark matter candidate(s) must be either cold or warm, stable or metastable, and 
collisionless, to the extent that the bulk of it ought not have 
substantial strong or electromagnetic charge. It is 
challenging to devise a dark-matter model which is consistent with all its 
known features, 
particularly if one hopes the candidate to be discoverable albeit not yet discovered.  
The space of possibilitites run the gamut in terms of possible masses 
and interaction cross sections with nucleons, and a sampling of the possibilities 
is shown in Fig.~\ref{fig:DMcand}. Many more possibilities exist, and the
field continues to evolve and produce new ones. We regard 
any model which can explain any of the observed dark-matter features
and/or answer any additional question unanswered in the standard model as 
well-motivated and hence of interest, though we consider
only a small fraction of the possibilities. 

We note in passing that contributions to the 
non-luminous halo of our own Milky Way galaxy 
could come from still more massive, compact objects. 
Such lumps could be of conventional matter 
and include faded white dwarfs, brown dwarfs, 
black holes, and neutron stars --- they are 
termed collectively ``Massive Astrophysical Compact Halo Objects'' or MACHOs. 
Their existence in the galactic halo has been probed primarily through searches for
gravitational microlensing events associated with the stars in the 
Large Magellenic Cloud~\cite{Alcock:2000ph,Alcock:1998fx,Tisserand:2006zx,Wyrzykowski:2011tr}.
Nonobservation of such events beyond expectation exclude MACHOs of mass ranging 
from $0.6\times 10^{-7}M_\odot < M < 15M_\odot$~\cite{Tisserand:2006zx,Wyrzykowski:2011tr} 
at 95\% CL, noting $M_\odot$ is the solar mass, 
as the dominant component of dark matter 
in our galactic halo~\cite{Alcock:2000ph,Alcock:1998fx,Tisserand:2006zx,Wyrzykowski:2011tr}.
Moreover, we should point out that the 
gravitational microlensing technique can 
be combined with Kepler results to search 
for primordial black hole dark matter 
in a black hole mass regime, such as that of planetary masses, 
not yet ruled out by other observations~\cite{Cieplak:2012dp}.

Although many dark-matter models possess candidates which can be
produced directly at colliders, 
we believe that the definite  
resolution of the dark-matter problem in terms of 
a candidate from particle physics will require detection of that
particle as a constituent of dark matter at the solar circle 
in a terrestrial experiment. Therefore we consider the notion of dark-matter
direct --- and indirect --- detection more generally 
before turning to specific dark-matter models. 
Candidates with weak-scale masses which couple to nuclei via weak neutral currents, or WIMPs, 
can be discovered through searches for 
nuclear recoil events from the aftermath of 
dark-matter--nucleus scattering~\cite{Drukier:1983gj,Goodman:1984dc}. 
We note, parenthetically, that candidates with 
sub-eV masses~\cite{Sikivie:1985yu,Asztalos:2009yp}, as 
well as warm-dark-matter candidates~\cite{Gardner:2006za,Gardner:2008yn}, can 
be detected directly through laser experiments. 
The interpretation of an anomalous-nuclear-recoil experiment in terms of WIMP parameters 
contains three ingredients: 
(i) the assumed dark-matter--nucleon interaction, (ii) the dark-matter 
number density and velocity distribution at the solar circle, 
and (iii) the computation of the relevant
nucleon matrix element of the appropriate current, or, more precisely, of the nuclear response 
it engenders. It has long been recognized that non-WIMP models can also be constrained
through such experiments, noting, {\it e.g.}, Ref.~\cite{Pospelov:2000bq}, and 
recently model-independent frameworks for (i) 
in the context of elastic dark-matter--nucleus scattering have been devised using 
effective-field-theory techniques~\cite{Fan:2010gt,Fitzpatrick:2012ix}. At
fixed order in an expansion in momentum transfer different interactions --- and 
nuclear responses --- are possible~\cite{Fitzpatrick:2012ix,Fitzpatrick:2012ib}.
This freedom is insufficient in itself to render inconsistent experimental results compatible
with each other~\cite{Fitzpatrick:2012ib} albeit differing astrophysical input (ii) 
also relaxes such tensions~\cite{Kelso:2011gd}. 

As for (ii), assumptions about 
the dark-matter 
mass density and velocity distribution 
are invariably necessary 
because, unfortunately, the local dark-matter distribution function is not known. 
Observational bounds 
on the dark-matter mass density $\rho_\chi$, {\it e.g.}, in our own solar system
are poor and 
exceed the 
estimates typically employed 
by orders of magnitude~\cite{Khriplovich:2006sq,Frere:2007pi}. 
Nevertheless, more direct-detection data and experiments should help constrain the
distribution function once
a signal is seen~\cite{Peter:2009ak,Fox:2010bu,Fox:2010bz,Peter:2011eu,Friedland:2012fa}.
In the canonical model
employed in the analysis of direct detection experiments, 
one assumes that the dark matter in the Milky Way resides in 
a non-rotating halo and that the velocity distribution 
$f(\bm{v})$ in that halo is that of an 
isothermal sphere~\cite{Lewin:1995rx}. The form of $f$ is thus that of a 
Maxwell-Boltzmann distribution centered on $v_0$ truncated by the Galactic escape 
speed $v_{\rm esc}$, noting $\rho_\chi=0.3\, \hbox{GeV/cm}^3$, $v_0=220\,\hbox{km/s}$, 
and 
$v_{\rm esc}=544\,\hbox{km/s}$ as employed, {\it e.g.}, in Ref.~\cite{Aprile:2012nq}. 
We note that known astrophysical effects prompt several refinements~\cite{Green:2002ht}. 
The formation of the Milky Way halo has also been studied in the context of
high-quality N-body simulations, which follow the accretion history of dark-matter clumps
over billions of years: early mergers yield a smooth halo, but more recent mergers
leave relic substructures, or subhalos~\cite{Diemand:2008in,Springel:2008cc}, 
and accretions of these clumps on the early galactic disk can 
bring additional complexities~\cite{Bruch:2008rx,Purcell:2009yp,Green:2010gw}. 
Tidal stripping of dark matter from subhalos
yields cold tidal streams and ``debris flows''~\cite{Lisanti:2011as,Kuhlen:2012fz}, 
so that simulations reveal a richly complex origin to dark matter at the solar circle,
which, in turn, can impact direct detection experiments~\cite{Stiff:2001dq}. 
Turning to observations, the existence of the Sagittarius stellar stream, produced
by the disruption and absorption of the Sagittarius dwarf galaxy by the Milky Way, 
 could impact 
the determination of the local dark-matter density
and its annual modulation; we refer to Ref.~\cite{Freese:2012xd} 
for a discussion of the possibilities. Recently, the role of the Sagittarius 
impact has been revisited
in detailed $N$-body simulations~\cite{Purcell:2011nf,Purcell:2012sh}, 
and a effect on local dark matter 
has been found~\cite{Purcell:2012sh}. The effect could also drive the 
vertical wave recently observed in the number counts of the local stars, signalling 
a departure from vertical equilibrium~\cite{Widrow:2012wu}, 
a connection itself supported by a numerical simulation~\cite{Gomez:2012rd}. 
Further observational 
studies of the local stars should help 
clarify the dark-matter 
distribution function at the Earth's location. In the next section we consider
the role of (iii) in the context of supersymmetric models. 

Dark matter can also be probed indirectly through the contribution 
of its decay and 
annihilation products to the budget of observed 
gamma and cosmic rays~\cite{Bergstrom:1997fj,Bergstrom:2000pn}. 
Generally the interpretation of such studies in favor of the presence of dark matter 
requires an understanding of the high-energy 
ejecta from conventional astrophysical sources~\cite{Bringmann:2012vr}. 
Two-body annihilation, however, yields a monoenergetic line and thus
is nominally background-free; the discovery of such lines would be 
experimentally challenging, though possible~\cite{Weniger:2012tx}.
The dark-matter
distribution, particularly the appearance of a 
dark disk, can also impact the annihilation rates~\cite{Bruch:2008rx,Bruch:2009rp}, 
as well as the  morphology of the
signal~\cite{Purcell:2009yp}. 
In recent years there has been much excitement over the discovery of 
excess gamma-ray or photon 
emission in various contexts, driven
by the interpretation of such as signals of dark matter, 
be it, {\it e.g.}, 
in the Galactic center~\cite{Hooper:2012sr,Abazajian:2012pn}, 
in bubbles extending from the Galactic 
center~\cite{Hooper:2013rwa}, or 
in the WMAP-Planck haze~\cite{Hooper:2010im}. In all the cases considered thus far, 
emission from conventional astrophysical sources, particularly milli-second 
pulsars~\cite{Abazajian:2012pn,Kaplinghat:2009ix}, 
could mimic the effects observed. 
It is worth noting that the angular distribution 
of the diffuse gamma-ray background can put constraints on, or 
even suggest a detection, of dark matter annihilation~\cite{SiegalGaskins:2008ge}.

We note in passing that indirect limits on dark-matter can also be realized
in terrestrial experiments, through collider studies~\cite{Bai:2010hh}, 
as well as through tests of the equivalence principle~\cite{Carroll:2008ub}. 
Torsion-balance experiments, both with and without spin-dependence, 
limit novel long-range forces~\cite{Adelberger:2009zz}, which can be
interpreted in a model-independent way~\cite{Dobrescu:2006au}, 
or as limits on particular models, such 
as axion models~\cite{Hoedl:2011zz}. 

We now review particular dark-matter models, starting with models with weak-scale
supersymmetry. 

\begin{figure}[t!]
\centering
\includegraphics[width=0.8\textwidth]{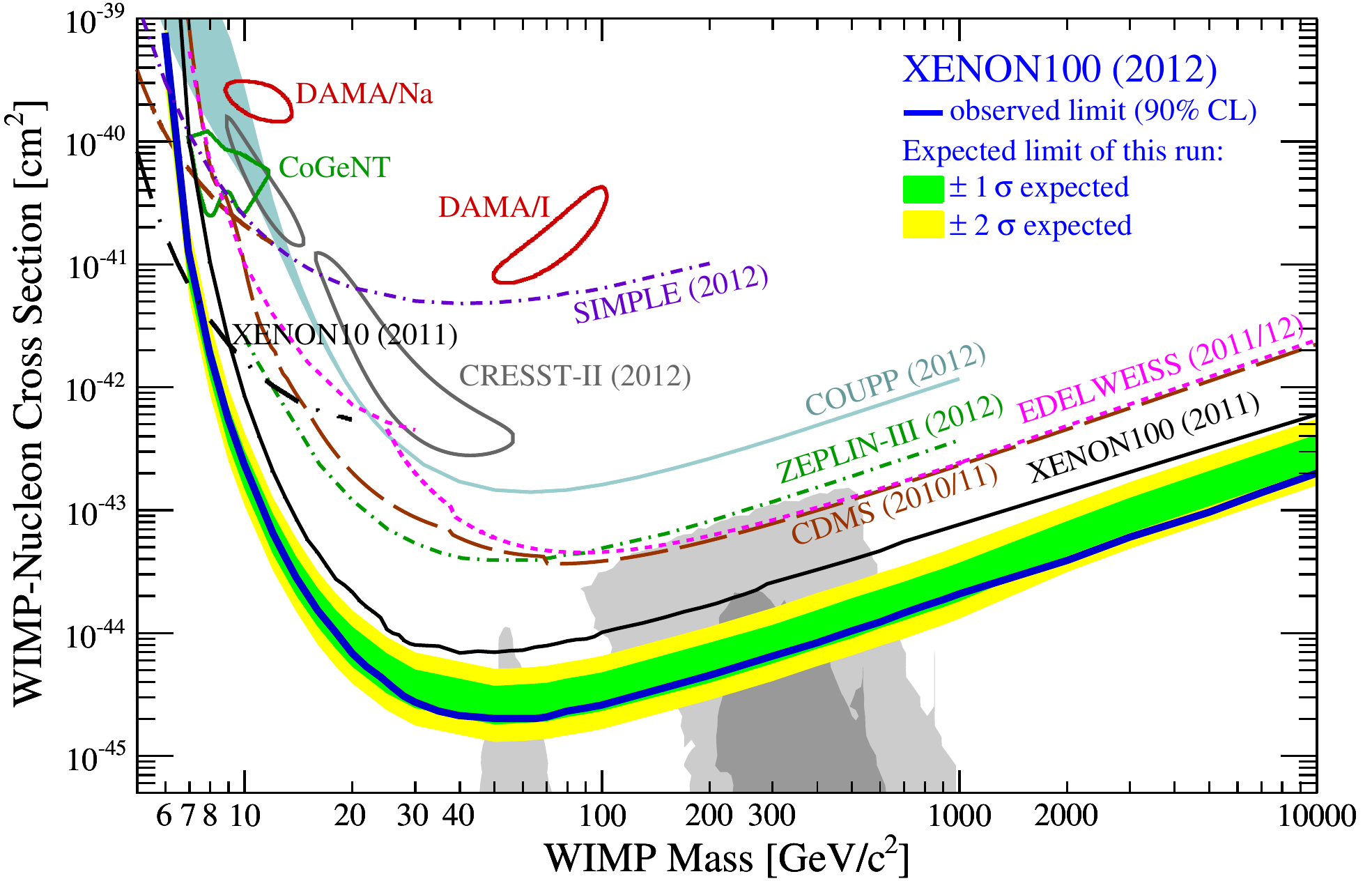}
\begin{minipage}[t]{16.5 cm}
\caption{
Constraints on the spin-independent WIMP-nucleon scattering cross section 
as a function of WIMP mass, figure
reprinted with permission from the online supplemental material of 
Ref.~\cite{Aprile:2012nq}.
The 90\% exclusion limit from the XENON100 (2012) experiment is 
shown in blue, as well as their expected sensitivities at 1$\sigma$ (green)
and 2$\sigma$ (yellow)~\cite{Aprile:2012nq}. Recent experimental results 
from the CDMS, CoGenT, COUPP, CRESST-II, EDELWEISS, SIMPLE, and ZEPLIN-II 
collaborations are also shown, as are the results from the DAMA, XENON10, 
and XENON100 (2011) experiments. The regions
at 1$\sigma$ and 2$\sigma$ preferred in particular (CMSSM) supersymmetric 
models are shown as well; we refer to Ref.~\cite{Aprile:2012nq} for all details. 
}
\label{fig:xenon100}
\end{minipage}
\end{figure}

\section{Supersymmetric models}

Models with weak-scale supersymmetry appeal for many reasons: 
(i) they can resolve the hierarchy problem, making the weak scale stable under
electroweak radiative corrections and thus technically ``natural,'' in a manner
consistent with precision electroweak measurements, 
(ii) they provide all the ingredients needed for successful electroweak 
baryogenesis, (iii) they can provide a suitable dark-matter candidate, and
(iv) they allow gauge coupling unification, at very high energy scales, to occur. 
A variety of theories fall under the aegis of weak-scale supersymmetry, 
and 
the minimally supersymmetric standard model (MSSM) is a particularly popular
variant. A particularly attractive feature of the MSSM is its ability
to draw together many issues in cosmology and particle 
physics~\cite{Cirigliano:2006dg}; it also has the ability to generate
electroweak symmertry breaking 
radiatively, in constrast to the standard model in which it is put in by hand. 
Nevertheless, the MSSM has flavor and CP problems in that the flavor-violating 
({\it e.g.}, an enhanced $B_s \to \mu^+\mu^-$ rate) 
and CP-violating ({\it e.g.}, a non-zero permanent electric dipole moment
of $^{199}${\rm Hg})  
effects it induces in low-energy observables have not been 
observed to occur, and direct searches for superpartner masses have also yielded
null results thus far. The null results are interconnected in that 
the low-energy problems can be remediated by simply making the superpartners more 
massive, note, {\it e.g.}, Ref.~\cite{McKeen:2013dma}, 
but this also makes the weak scale less natural. 
We refer to  Ref.~\cite{Feng_rev:2013} for 
a review of the current status of naturalness and supersymmetry. 

It has been long thought that the dominant component of dark-matter is a WIMP, 
and the MSSM offers a candidate in the form of the 
neutralino~\cite{Jungman:1995df}.\footnote{Other models, such 
as models with universal extra dimensions~\cite{Servant:2002aq,Cheng:2002ej} and 
branon models~\cite{Cembranos:2003mr}, also offer WIMP dark-matter candidates.} 
It is worth noting that the dark-matter 
stability 
requirement is challenging: it requires the imposition of an additional 
discrete symmetry, termed ``$R$-parity''~\cite{Goldberg:1983nd,Ellis:1983ew}. 
As we have noted, an appealing aspect of such 
a dark-matter candidate is that it is amenable to direct detection through
searches for anomalous elastic scattering events from nuclei, 
where we refer to Fig.~\ref{fig:xenon100} for a succinct summary of the current 
results. The exclusion limits have nearly reached the 
$10^{-45}\,{\rm cm}^2$ scale, which is $10^{-9}\,{\rm pb}$ --- a quick check
of Fig.~\ref{fig:DMcand} shows that we have eliminated at best half of the expected
WIMP parameter space.\footnote{We refer 
to Refs.~\cite{GeringerSameth:2011iw,GeringerSameth:2012sr}
for WIMP exclusion limits from indirect detection experiments.} 
The loci of points in shaded grey indicate the phenomenologically
acceptable parameter space 
associated with a particular variant of the MSSM with far fewer free
parameters: the CMSSM. As one observes, 
it is ``easy'' to build models without an appreciable 
direct detection signature, and that is not the least of it. 
We do not know the mechanism of supersymmetry breaking, and the 
 MSSM reflects that ignorance through the appearance of free parameters
which characterize  ``soft'' supersymmetry breaking --- there are 
an unwieldly number, some 124 in all, and there are also neutrino
parameters to consider. In making assumptions 
to limit the parameter space, we may fail to appreciate the scope 
of possibilities within the theory~\cite{Berger:2008cq}. For example, 
it is possible to arrange neutralinos which are much lighter than the 
weak scale in mass, and, indeed, they are not massless simply because 
cosmology bounds their mass from 
below\cite{Dreiner:2007fw,Dreiner:2009ic,Profumo:2008yg}.  
It is also possible to arrange supersymmetric models 
with multi-component dark matter, such as a WIMP with a particle akin to 
a sterile neutrino~\cite{Zurek:2008qg}. Moreover, it is possible to arrange 
supersymmetric models in which the lightest supersymmetric particle is a 
gravitino, through mechanisms in which flavor physics problems are absent. 
We note that very light gravitino candidates can connect to $N_{\rm eff}$, 
but only if they are not thermal relics~\cite{Feng:2010ij}. 
The sweep of possibilities in regards to supersymmetric dark matter
is vast, and 
it may prove immensely challenging in this context 
to falsify supersymmetry 
as a phenomenological construct.

The direct detection of dark matter entwines astro-, particle, and nuclear physics, 
and as a final topic we examine, recalling (iii) of the previous
section, the computation of the hadronic matrix elements 
germane to WIMP-nucleon scattering. 
In nuclear physics, the 
decipherment of the flavor and spin structure of the proton and neutron 
is a topic of ongoing intense interest, and it 
also has broad 
implications for the search for physics BSM~\cite{Gardner:2010zf,Brodsky:2012zza}. 
In our current context, the 
strange-quark structure of the nucleon impacts the interpretation of
experiments which hunt for WIMP dark matter
in that it impacts the mapping of the loci of supersymmetric parameter space
to the exclusion plot of WIMP mass versus the WIMP-nucleon cross section, 
as per Fig.~\ref{fig:xenon100}. 
The spin-independent 
neutralino-nucleon 
cross section is particularly sensitive to the strange scalar 
density, namely, the value of 
$y=2 \langle N | \bar s s | N \rangle/
\langle N | \bar u u + \bar d d | N \rangle$~\cite{Ellis:2008hf}, because the 
neutralino coupling increases with quark mass; accordingly, the 
spin-dependent neutralino-nucleon cross section is sensitive to the 
strange quark axial vector matrix element, a topic of intense
interest for many years in nuclear physics~\cite{Musolf:1993tb}. Here we focus 
on the spin-independent case in order to interpret Fig.2.
Earlier studies relate $y$ to the $\pi N$ sigma term $\Sigma_{\pi N}$ via 
$y=1- \sigma_0/\Sigma_{\pi N}$ for fixed 
$\sigma_0 \equiv m_l 
\langle N | \bar u u 
+ \bar d d - 2 \bar s s | N \rangle$~\cite{Ellis:2008hf}, with 
$m_l\equiv(m_u+m_d)/2$, 
so that the 
predicted neutralino-nucleon cross section would seem to depend strongly 
on the phenomenological value of the $\Sigma_{\pi N}$ term~\cite{Giedt:2009mr}, 
for which there is a spread of determined values~\cite{Young:2013nn}. 
However, 
$m_s \langle N | \bar s s | N \rangle$ and 
$\Sigma_{\pi N} \equiv m_l\langle N | \bar u u + \bar d d | N \rangle$
can be computed directly in lattice QCD, via different techniques, 
and the final neutralino-nucleon cross
section is not nearly as sensitive to 
$\Sigma_{\pi N}$ as earlier thought~\cite{Giedt:2009mr}. 
Several lattice QCD groups have addressed this problem and new results
continue to emerge~\cite{Junnarkar:2013ac}; we 
refer to Ref.~\cite{Young:2013nn} for a recent review. 
The outcome of this body of work is that 
the spin-independent
WIMP-nucleon cross section can be predicted to much better precision than 
previously thought, though the cross section tends to be smaller 
than that previously assumed~\cite{Giedt:2009mr}, 
diminishing the new physics reach of 
a particular WIMP direct detection experiment. 
The allowed CMSSM parameter space of Fig.~\ref{fig:xenon100} does not seem 
to incorporate these updates~\cite{Young:2013nn}, so that the constraints
on the CMSSM parameter space may not be as strong as had been 
thought~\cite{Aprile:2012nq}. 
Heavier quark flavors can also play a significant role in 
 mediating the gluon coupling to the Higgs, and hence to the neutralino, and the 
leading contribution in the heavy-quark limit is 
well-known~\cite{Shifman:1978zn,Jungman:1995df} --- and this treatment 
should describe elastic scattering sufficiently well. 
Recently, interpreting the conflicting tangle of possible dark-matter 
signatures has led to the suggestion of composite
dark-matter candidates~\cite{Finkbeiner:2007kk,Finkbeiner:2009mi}; 
here the intrinsic heavy quarks could play a more interesting 
role in mediating transitions to excited dark-matter states
in scattering experiments~\cite{Brodsky:2012zza}. 
We note in passing that 
WIMP-nucleon~\cite{Hill:2011be} and WIMP-nucleus ~\cite{Cirigliano:2012pq} 
scattering have also been studied in effective-field theory. 

Developing experimental and observational tensions with the predictions
of supersymmetric models encourage broader thinking in regards to 
the composition of dark-matter, and we consider some well-motivated
alternatives in the sections to follow. 

\section{Hidden Sector Models} 

If dark matter is not made of WIMPs, its stability need not be guaranteed 
by a discrete symmetry, and its relic density need not be fixed by thermal 
freezeout. These features could potentially be explained in very different ways. 
What mechanisms, then, could be operative?

\begin{itemize} 

\item Its stability could be guaranteed by a hidden gauge symmetry. 

\item Its relic density could be related to the baryon asymmetry. 
If so, dark matter ought be {\it asymmetric}. 

\end{itemize}

In this section we begin with the first possibility: models which
possess a hidden gauge symmetry. We note that models which 
simultaneously explain dark matter and the baryon asymmetry 
invariably possess hidden gauge symmetries as well~\cite{hoorabi}, 
though we reserve discussion of such models for the moment.

The study of hidden-sector models has gained impetus from hints of new physics
in indirect detection experiments. The PAMELA experiment, {\it e.g.}, 
can detect charged particles, {\it i.e.}, $e^-$, $e^+$, $p$, and $\bar p$, from space,
and observes excess events in the ratio of $e^+$ to $e^-$ final states but no anomalies 
in the ratio of $\bar p$ to  $p$ 
final states~\cite{Adriani:2008zq,Adriani:2008zr,Adriani:2010ib}. 
Such a pattern, if from dark matter, would not easily arise in a
supersymmetric model; rather, these results can be taken to suggest that dark matter
has preferential couplings to leptons~\cite{Fox:2008kb}. 
Taken in concert with the 
results from the  DAMA experiment, the results 
promote the notion that the dark-matter candidate has 
internal structure~\cite{Finkbeiner:2009mi}, which is also 
suggestive of a hidden gauge symmetry.  The cosmic ray excess in leptons
can also be explained if dark matter annihilates into an intermediate state lighter
than the proton in mass~\cite{Cholis:2008qq}, which can be arranged 
in models with a hidden-sector 
gauge symmetry~\cite{ArkaniHamed:2008qn,Pospelov:2008jd,Finkbeiner:2008qu}. 
The excesses found by PAMELA are supported other experiments, such as 
FERMI~\cite{FermiLAT:2011ab}, though 
an explanation may ultimately be found  to derive from 
conventional astrophysical sources. We note that the AMS experiment
has the capacity to study the cosmic ray spectrum at 
yet higher energies, where presumably 
conventional sources play less of a role. 
We regard the existing results as evocative of the possibilities, and a hidden
sector operating under a U(1) gauge symmetry is merely the simplest 
among them. Interestingly the narrow value of the determined 
Higgs mass and a possible vacuum stability problem 
can also point to the existence of new U(1) 
interactions~\cite{Chao:2012mx}, 
and this possibility is under evaluation~\cite{Chao:2012mx,Bulava:2012rb}. 
In what follows we organize our discussion in terms of the manner 
in which hidden degrees of freedom could connect to the particles
of the standard model, for that predicates their detectability. 
Note that models which couple to the hidden sector through a Higgs
portal have also been considered~\cite{Patt:2006fw,Barger:2007im,FileviezPerez:2008bj,Barger:2008jx,Gonderinger:2009jp,He:2009yd,Cheung:2012nb,Gonderinger:2012rd}, though we do not discuss them. 

{\it Hermetic Models} are those in which the dark sector is blind to 
all standard model gauge interactions. Yet even in such cases observational 
constraints can be made. Suppose, {\it e.g.}, an exact U(1) symmetry operates 
in the hidden sector~\cite{Feng:2008mu,Feng:2009mn} --- a dark electromagnetism. 
Dark matter would then carry a hidden charge 
and be stable just as the electron is stable. An explicit example of such 
a model with a hidden MSSM-like sector is considered in Ref.~\cite{Feng:2009mn}, 
so that the putative dark-matter has both hidden weak and electromagnetic interactions. 
Such a model can 
have the right dark-matter relic density and yield cold dark matter, and can 
generally be cosmologically indistinguishable from usual WIMP models. 
Dark matter in this model 
is significantly self-interacting, however, 
with long-range forces. This makes it subject to observational
constraints, most notably the self-interaction
constraint from observations of the Bullet Cluster and the observed
ellipticity of dark-matter halos, as kinetic energy transfer through 
dark-matter elastic scattering
would tend to isotropize the mass distribution~\cite{Feng:2009mn,Ackerman:2008gi}. 
Such considerations give 
constraints on the hidden fine-structure constant $\alpha_\chi$ as
a function of candidate mass $M_\chi$, yielding, 
{\it e.g.},  
$\alpha_\chi < 10^{-7}$ for $M_\chi \sim 1$ GeV~\cite{Feng:2009mn}. 
    Self-interaction constraints from halo morphology have recently 
    been revisited, and some argue that {\it evidence} exists 
    for self-interacting dark matter~\cite{Rocha:2012jg,Peter:2012jh}.

\begin{figure}[t!]
\centering
\includegraphics[width=0.6\textwidth]{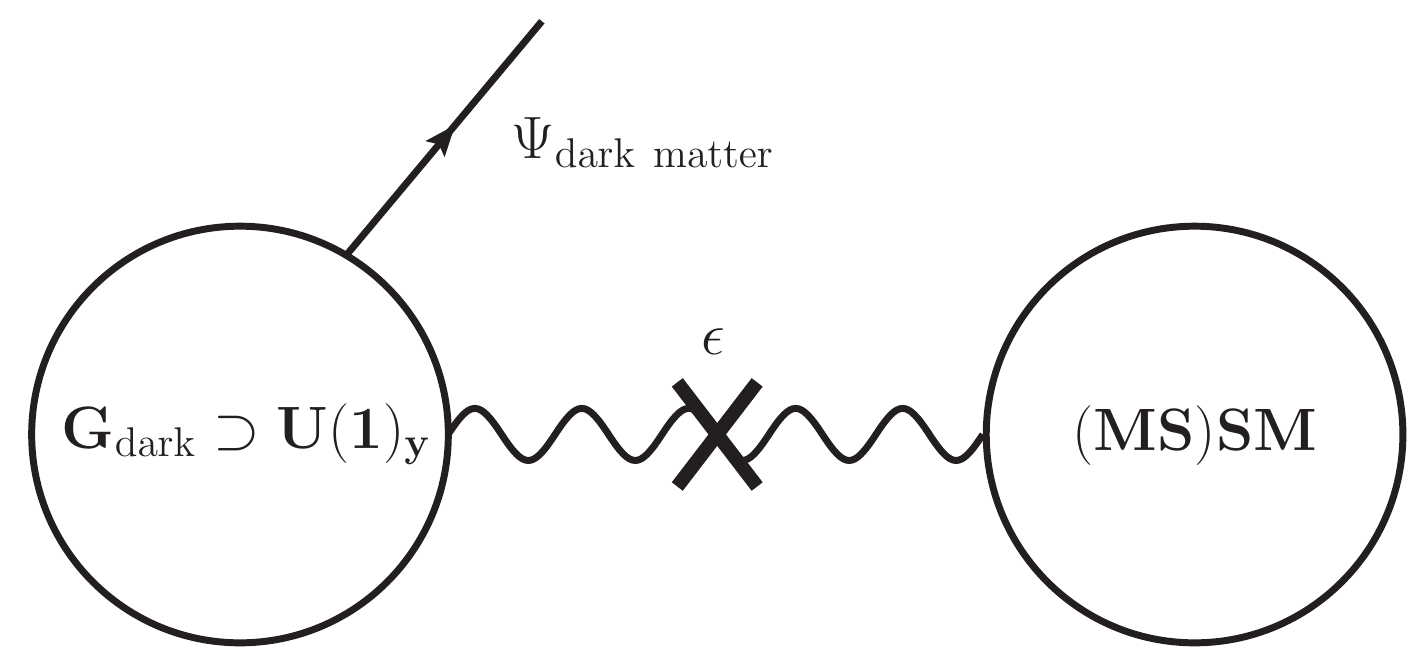}
\begin{minipage}[t]{16.5 cm}
\caption{
A non-abelian dark sector can contain an abelian ideal, 
 which permits
kinetic mixing with the gauge bosons of the standard model, or MSSM, through a 
marginal 
operator. 
Illustration reprinted with permission from Ref.~\cite{Baumgart:2009tn}.
}
\label{fig:connect}
\end{minipage}
\end{figure}

{\em Models with Abelian Connectors} are inspired, in part, by the astrophysical 
anomalies we have described, though broader possibilities also exist, 
which are not tied to such signals. {\it E.g.}, a
hidden sector electromagnetism with a ``paraphoton,'' which mixes with the photon
through kinetic mixing, is an idea of long standing~\cite{Holdom:1985ag,Holdom:1986eq}. 
It is also amenable to experimental test, perhaps most notably through searches
for ``light shining through walls''~\cite{Jaeckel:2010ni} --- tests
which are also possible at the FEL at JLab~\cite{Afanasev:2008fv}.
This also has consequences for dark matter, in that 
if the hidden gauge mediator is massless, although this 
is not a necessary condition~\cite{Feldman:2007wj}, dark matter 
can have a {\em millicharge}~\cite{Holdom:1986eq}. 
Consequently these ideas are also tested through millicharged particle
searches. Interestingly, 
if we determine that dark matter has a nonzero 
millicharge $\varepsilon e$, no matter how small, we establish 
that dark matter is stable 
by dint of a gauge symmetry --- 
it cannot decay and conserve its electric charge. We refer to 
Ref.~\cite{Davidson:2000hf} for a comprehensive review. We note that 
a direct limit on the dark-matter (milli)electric-charge-to-mass ratio can 
be realized from the time delay 
of radio afterglows from gamma-ray 
bursts, yielding $|\varepsilon|/M < 1\times 10^{-5}\,{\rm eV}^{-1}$ at
95\% CL~\cite{Gardner:2009et}. This limit can be enormously bettered if
``prompt'' radio afterglows can be detected at extremely low frequencies, 
such as possible at 
LOFAR~\cite{Morganti:2011hg}. Millicharged matter limits also follow from 
the nonobservation of the effects of millicharged 
particle production, and these typically prove to yield the best limits. 
 The strongest such bound from laboratory experiments 
is $\vert \varepsilon \vert <  3-4 \times 10^{-7}$ 
for $M \le 0.05$ eV~\cite{Ahlers:2007qf}, so that for 
$M\sim 0.05\,\hbox{eV}$ the limits are crudely comparable. 
Indirect limits also emerge from stellar 
evolution constraints, 
for which the strongest is $|\varepsilon| < 2\times 10^{-14}$ for 
$M< 5\,\hbox{keV}$~\cite{Davidson:2000hf}, 
as well as from the manner in which numerical 
simulations of galactic structure confront 
observations~\cite{Gradwohl:1992ue,Feng:2009mn,Ackerman:2008gi}. 
Such limits can be evaded; in some models, the 
dynamics which gives rise to millicharged matter are not operative at 
stellar temperatures~\cite{Masso:2006gc}; other 
models evade the galactic structure constraints~\cite{Kusenko:2001vu}. 

We now turn to the models spurred by the 
intriguing
astrophysical anomalies we have noted. 
The visible and hidden sectors are connected through the 
kinetic mixing of the gauge bosons of 
their respective U(1) symmetries, notably through a 
standard model hypercharge U(1)$_Y$ 
portal~\cite{Holdom:1985ag,ArkaniHamed:2008qn,Baumgart:2009tn,Essig:2009nc,Bjorken:2009mm}.  
We refer to Fig.~\ref{fig:connect} for an illustration; it is worth noting that 
${\rm G}_{\rm dark}$ can be a rich choice; the hidden sector could be, e.g., 
MSSM-like, as in the model of Ref.~\cite{Feng:2009mn}. 
Constraints on long-range 
interactions between dark-matter particles are 
sufficiently severe~\cite{Spergel:1999mh,Ackerman:2008gi,Feng:2009mn} that 
in the  models we consider the dark gauge symmetries are also 
broken through a dark Higgs sector, note, {\it e.g.}, 
Ref.~\cite{Baumgart:2009tn}, giving 
a mass to the hidden gauge boson --- and dark matter no longer has a millicharge. 
If we suppose 
$A^\prime$ is the gauge field of a massive dark U(1)$^\prime$ gauge group, then
the standard model Lagrangian ${\cal L}_{SM}$ is enlarged to~\cite{Bjorken:2009mm} 
\begin{equation}
{\cal L}={\cal L}_{SM} + \frac{\epsilon_Y}{2} F^{Y,\mu \nu} F^\prime_{\mu \nu}
- \frac{1}{4} F^{\prime,\mu \nu} F^\prime_{\mu \nu} + 
m_{A^\prime}^2 A^{\prime\,\mu} A^{\prime}_\mu \,,
\end{equation}
where, {\it e.g.}, $F^\prime_{\mu \nu} \equiv \partial_\mu 
A^\prime_\nu - \partial_\nu A^\prime_\mu$. 
Moving from the gauge to mass eigenstate basis, we can redefine
the photon and paraphoton fields so that the kinetic mixing term disappears, namely 
via $A_\mu \to \tilde A_\mu = A_\mu^\prime - \epsilon A_\mu^\prime$, 
with $\epsilon\equiv \varepsilon_Y \cos \theta_W$; and we discover that 
$A^\prime_\mu$ couples slightly to the electromagnetic current. 
It couples to the $Z_\mu$ as well, but this effect is suppressed by a
factor of $m_{A^\prime}^2/m_Z^2$~\cite{Baumgart:2009tn,Bjorken:2009mm}. 
Since the kinetic mixing term is of mass dimension 4 it 
can be thought of as a UV boundary condition; equivalently, one notes 
there is no energy at which it must cease to be valid. 
If heavy particles exist which are charged under both U(1) groups, 
an estimate of $\epsilon$ follows from the computation of the associated 
loop-induced effect, indicating that $\epsilon$ is no greater 
than ${\cal O}(10^{-2})$; moreover, 
if the U(1) symmetry-breaking effects 
are connected to the weak scale, such effects reveal that the $A^\prime$ 
can range from the MeV- to GeV-scale in mass.

\begin{figure}[t!]
\centering
\includegraphics[width=0.6\textwidth]{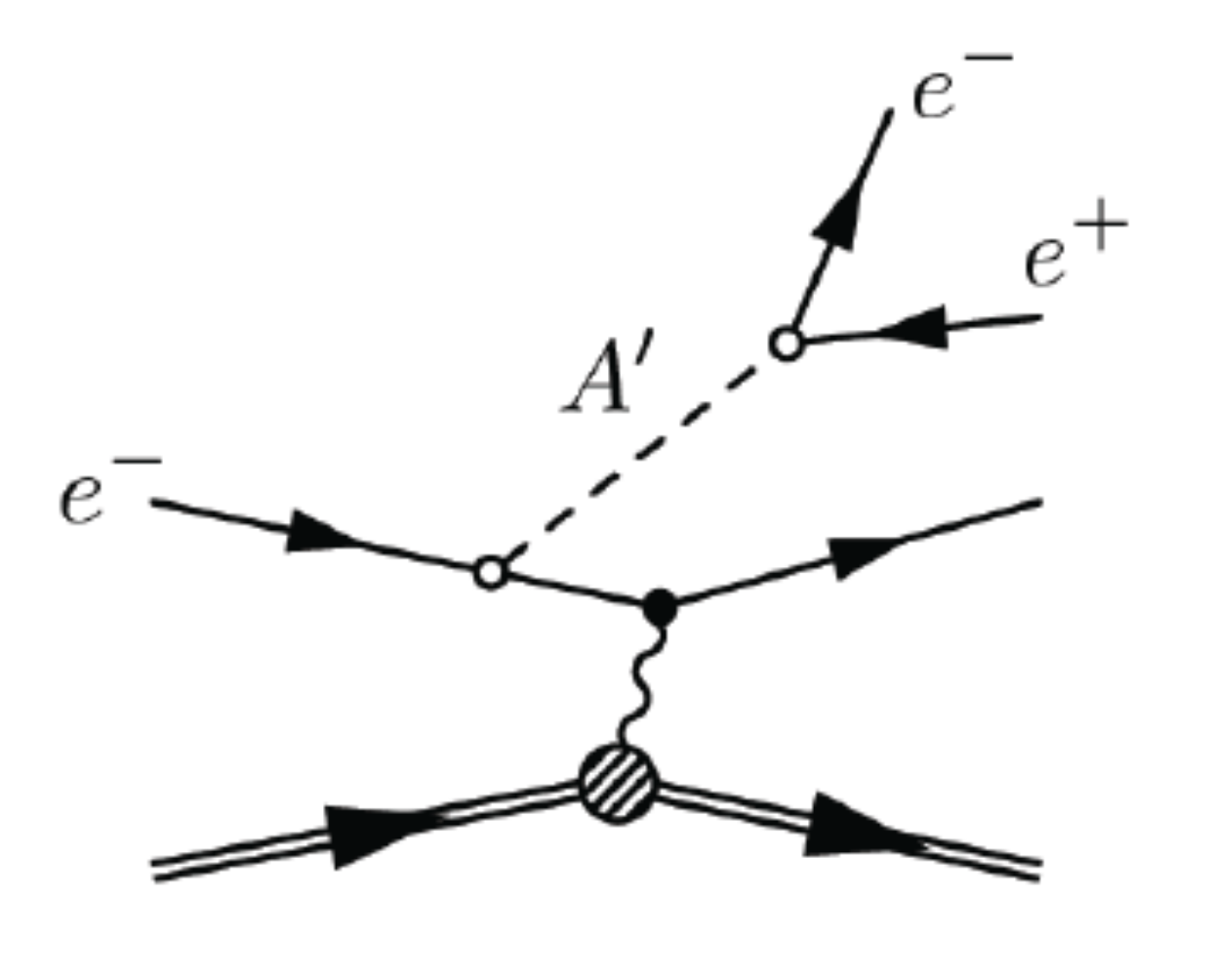}
\begin{minipage}[t]{16.5 cm}
\caption{
An illustration of the manner in which a hidden gauge boson $A^{\prime\,\mu}$
can participate in a fixed-target experiment, 
figure reprinted with permission from Ref.~\cite{McKeown:2011yj}. 
}
\label{fig:rxn_mech}
\end{minipage}
\end{figure}

\begin{figure}[t!]
\centering
\includegraphics[width=0.7\textwidth]{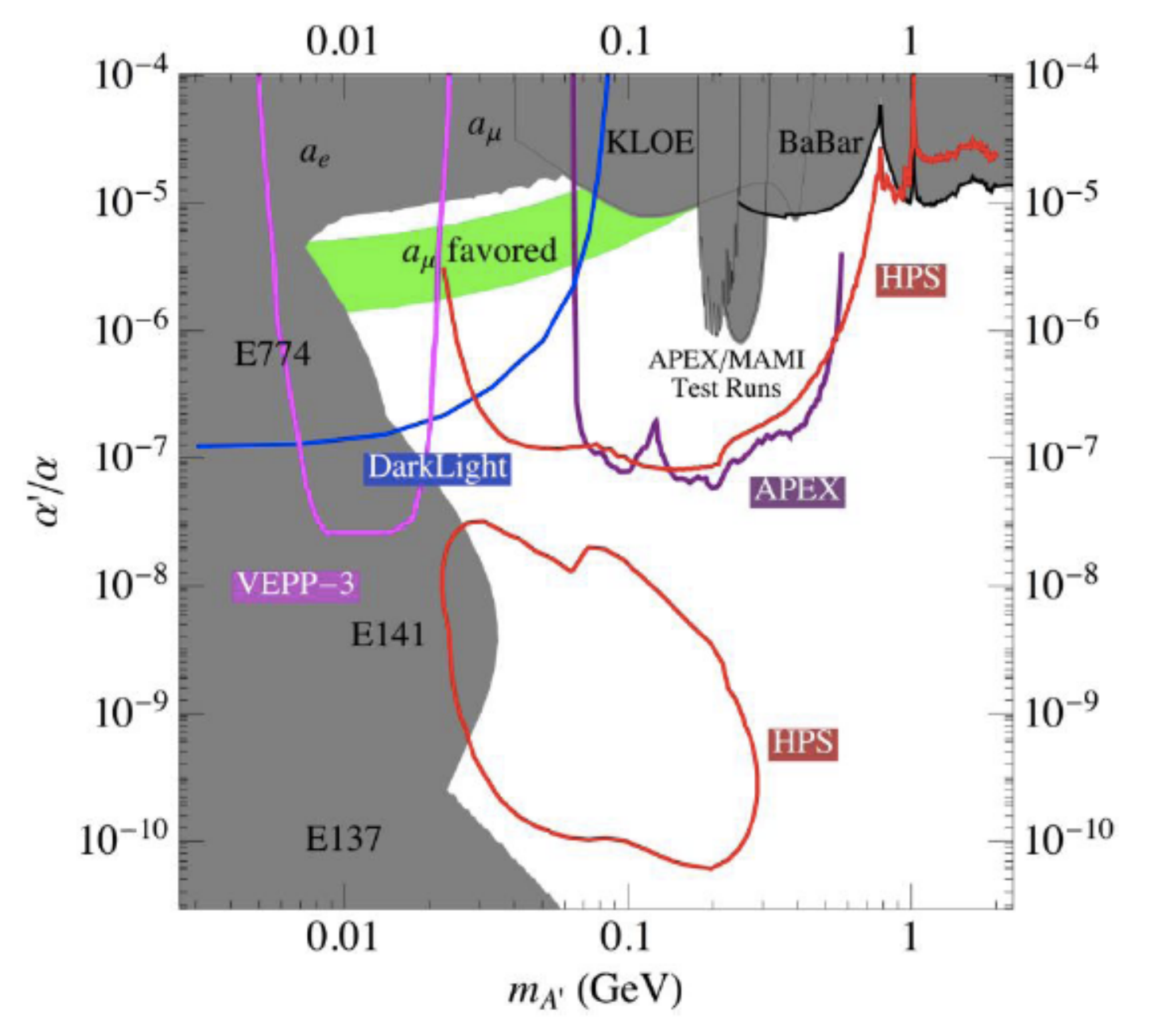}
\begin{minipage}[t]{16.5 cm}
\caption{
Constraints on the hidden-sector fine-structure constant $\alpha^\prime$ 
as a function of the $A^\prime$ mass $M_{A^\prime}$, 
figure reprinted with permission from Ref.~\cite{Dudek:2012vr}.
Shaded regions show the limits at 90\% CL 
which emerge from the beam dump experiments E137, E141, E774, from KLOE and BaBar, and
from the test run results reported by APEX (JLAB) and A1 (MAMI). The limits from 
the muon and electron anomalous magnetic moments, $a_\mu$ and $a_e$, are shown as
well. The shaded band labelled ``$a_\mu$ favored'' 
shows the regio in which the $A^\prime$ can resolve the observed
discrepancy in $g-2$ of the muon at 90\% CL~\cite{Pospelov:2008zw}. 
The improved thtical computation of $a_e$~\cite{Aoyama:2012wj} sharpens
the interpretaton of its measurement~\cite{Hanneke:2008tm,Bouchendira:2010es}
and removes some of the ``$a_\mu$ favored'' region, see Ref.~\cite{Davoudiasl:2012ig}
for an illustration. 
Projected sensitivities of the APEX, DarkLight, HPS, and VEPP3 experiments are 
shown as well. 
We refer to Ref.~\cite{Dudek:2012vr} for all details. 
}
\label{fig:jlab}
\end{minipage}
\end{figure}

An appealing feature of the $A^\prime$ is that it 
can be discovered in fixed-target experiments at nuclear-physics facilities; 
we note an illustration of 
how it may do so in Fig.~\ref{fig:rxn_mech}. 
Constraints on the $A^\prime$ follow from searches for fractionally charged
particles in beam dump experiments, from studies of meson decays, and 
from measurements of the anomalous magnetic moment of the electric and
muon~\cite{Fayet:2007ua,Pospelov:2008zw} --- we refer to 
Ref.~\cite{Bjorken:2009mm} for a comprehensive study. 
Fig.~\ref{fig:jlab} illustrates 
the existing limits 
on the mass of the $A^\prime$ and 
its hidden fine-structure constant $\alpha^\prime \equiv \varepsilon^2/4\pi$, 
as well as 
the constraints which can emerge from future experimental 
studies at JLab, MAMI, and Novosibirsk. 

{\em Models with non-Abelian Connectors} are those 
in which the connection between the hidden and visible sector is through 
a non-Abelian portal. 
The notion of a hidden sector of strongly coupled matter 
is of some standing~\cite{Okun:2006eb,Berezhiani:2003xm}, and has more recently been 
discussed in the context of models which provide a 
common origin to baryons and dark matter~\cite{Nussinov:1985xr,Barr:1990ca}, 
though the mechanism need not be realized through strong 
dynamics~\cite{Kaplan:1991ah,zurek} ---
we note Ref.~\cite{hoorabi} for a recent review. 
We consider a non-Abelian portal~\cite{Gardner:2013aiw}, mediated, {\it e.g.}, 
by heavy scalars $\Phi$ which transform under the adjoint representation of the 
non-Abelian group SU(3); 
such an interaction can also be realized through kinetic mixing, generalizing from 
Ref.~\cite{Baumgart:2009tn}, through 
${\rm tr}(\Phi F_{\mu\nu}){\rm tr}(\tilde\Phi {\tilde F}^{\mu\nu})$, as well as 
$\epsilon^{\mu\nu\rho\sigma} {\rm tr}(\Phi F_{\mu\nu})
{\rm tr}(\tilde\Phi {\tilde F}_{\rho\sigma})$, where 
$F^{a\,\mu\nu}$ is the standard model SU(3)$_c$ field strength, and 
$\tilde{\Phi}^a$ and $\tilde{F}^{a\,\mu\nu}$ are 
fields and field strengths of a hidden strongly-coupled sector, nominally 
based on SU(3)$_{\tilde c}$. We anticipate 
that the dark matter candidate is a composite particle and a color singlet, so
that there are no dark long-range forces to negate. 
The connector is not a marginal operator, so that the model does not
have a clear UV completion --- it represents an effective theory. 
We note that 
the appearance of QCD-like couplings should make it more important in the 
infrared. At low energies the physics of confinement prompt the use of
the hidden-local-symmetry model of QCD: the $\rho$ meson emerges as its
dynamical gauge boson. Thus the coupling of visible and hidden sectors 
can be modelled in terms of a
kinetic mixing model with two massive gauge bosons --- a $\rho$ and $\rho^\prime$,
both with isospin 1. The appearance of the $\rho^\prime$ is hidden under
hadronization uncertainties, but one can hope to detect its presence
through its possible CP-violating effects, as through the study of 
pseudo-T-odd momentum correlations in radiative $\beta$ decay of neutrons
and nuclei~\cite{Gardner:2013aiw,Gardner:2012rp}, 
which can be studied at existing and future radioactive beam 
facilities. More generally we can think of the 
$\rho'$ as a mediator in realizing 
a difference in the radiative $n$ and $\bar n$ $\beta$ decay rates, 
motivating a measurement of the $\bar n$ lifetime. If there were 
a U(1)$_Y$ portal as well, we would have a composite dark-matter candidate
with a magnetic moment, which could be detected through 
its elastic scattering from nuclei~\cite{Bagnasco:1993st} 
or through a laser experiment, such as 
through detection of a magnetic Faraday effect~\cite{Gardner:2008yn}. 

These discussions naturally lead us to our final topic: of {\it asymmetric}
dark matter, in which baryons and dark matter share a common origin. 
A key take-away message of the observations is that 
the baryonic rest mass contribution to closure 
is roughly $20\%$ of the overall 
dark matter contribution. 
This is not a small fraction, and its 
magnitude begs the question of why, {\it e.g.}, 
the baryon and CDM contributions to closure are so close in size. 
In these models dark matter is a fermion, and thus it possesses its own particle 
{\it asymmetry}, which can discovered through a measurement of a non-zero
magnetic Faraday effect~\cite{Gardner:2006za,Gardner:2008yn}. 
For detailed models we 
simply note the review of Ref.~\cite{hoorabi}. From the viewpoint of low-energy
physics, it is worth noting that 
interesting features such as dark-matter particle-antiparticle 
oscillations can 
appear in such models~\cite{Tulin:2012re}.

\section{Sterile Neutrinos}

The advances in experimental neutrino physics in the last decade have been unprecedented.
The laboratory measurements have given us the neutrino mass-squared differences and three ($\theta_{ 1 2}$, $\theta_{ 2 3}$, $\theta_{1 3}$) of the four parameters which characterize the unitary transformation between neutrino energy states (\lq\lq mass\rq\rq\ states) and the weak interaction eigenstates (flavor states) in vacuum. 
All we are missing is the fourth parameter, the CP-violating phase, 
though we note that there are potentially also additional Majorana CP phases. 
Of course we are also ignorant of the actual vacuum neutrino mass eigenvalues
and the way these are ordered, {\it i.e.}, the neutrino mass hierarchy. 

However, even absent this missing information there are two overwhelming standout features of the experimental results: the neutrinos have rest masses; and these are very small compared to the rest masses of the other elementary particles in their respective families. Once an active neutrino has a nonzero mass it could flip its spin from left- to right-handed. Right-handed Dirac neutrinos and left-handed Dirac antineutrinos do not interact via the weak force. These particles really would be sterile. However, models can be made where these particles mix in vacuum with ordinary active neutrinos which can be either Majorana or Dirac in character. The designation \lq\lq sterile,\rq\rq\ was inspired by how a massless Dirac right-handed neutrino or left-handed antineutrino would behave. But by sterile neutrino here we shall mean any chargeless spin-1/2 fermion which has sufficiently sub-weak interaction coupling that it is not ruled out by the $Z^0$ width limits, {\it e.g.}, from LEP.  

The many attempts to explain the disparity in rest mass scales between the known neutrinos and the other elementary particles mostly revolve around \lq\lq see-saw\rq\rq\ models \cite{Minkowski:1977zr,Yanagida:79,Gell-Mann:80,Glashow:1980,Mohapatra:1980mz} . In these schemes it is posited that the product of the mass scale associated with the known active neutrinos and the mass scale of some \lq\lq sterile\rq\rq\ species is the square of an extremely large mass-scale, such as 
the unification scale, for example. Very heavy sterile neutrinos then imply very light active neutrinos, \lq\lq explaining\rq\rq\ why active neutrinos are so light and why sterile states do not show up in accelerator experiments and in astrophysical settings 
such as core collapse supernovae and BBN.

We can conclude only that the existence of sterile neutrinos is at least plausible.  Ref.~\cite{Kusenko:2009lr} provides a comprehensive review of sterile neutrinos, evidence for these particles and constraints on them, and their possible effects in astrophysical settings ranging from the early universe to compact objects. 

Disturbingly, though the LEP results require only three active neutrinos with standard weak interactions, there is no limit on the number of sterile neutrinos. Furthermore, there are no compelling arguments for what the rest masses of sterile neutrinos should be. In fact, there are credible, if not persuasive, arguments for sterile neutrino rest masses ranging from the sub-eV scale to the unification scale (see for example Refs.~\cite{de-Gouvea:2005fk,de-Gouvea:2007lr}). 
The see-saw mechanism usually is based on invoking ${\cal O}(1)$ 
Yukawa couplings to the Higgs and, of course, on 
heavy right-handed neutrino masses.  Interestingly, however, the split seesaw mechanism \cite{Kusenko:2010qy} can reconcile active neutrino masses with a relatively light sterile neutrino, {\it e.g.}, one with a mass well below the electroweak scale.  
Such a sterile neutrino is a natural dark matter candidate. 

The idea of an electroweak singlet (sterile neutrino) as a dark matter candidate has a long history at this point \cite{Dodelson:1994rt,Shi:1999lq,Abazajian:2001lr,Dolgov:2002ve,Abazajian:2002bh,Asaka:2005fj,Abazajian:2006qf,Shaposhnikov:2006fj,Boyanovsky:2007lr,Boyanovsky:2007fk,Shaposhnikov:2007qy,Gorbunov:2007uq,Kishimoto:2008pd,Laine:2008kx,Petraki:2008yq,Petraki:2008vn}. 
The sterile neutrino dark matter candidates in many models have rest 
masses of $\sim 1\, {\rm keV}$. 
In most of these models the sterile neutrinos mix in vacuum with active neutrino species. This gives the sterile neutrino an effective interaction in vacuum and in a 
medium ({\it e.g.}, in the early universe). These interactions imply that sterile neutrinos can have effects in astrophysical environments that can lead to constraints.

There are many examples of such effects and constraints derived from them. Sterile neutrinos have also been studied as a potential source of early re-ionization in the dark ages in the adolescent universe \cite{Biermann:2006mz,Mapelli:2006gf,Stasielak:2007ly}. They have been invoked to produce large pulsar kicks \cite{Kusenko:1999rt,Barkovich:2004ul,Fuller:2003vn,Loveridge:2004fr,Kishimoto:2011fk}, and in baryogenesis scenarios \cite{Akhmedov:1998ys,Asaka:2005fr}. Interestingly, they also can play havoc in the core collapse supernova environment \cite{Hidaka:2006yq, Fryer:2006rt,Hidaka:2007kx,Fuller:2009uq}. 

At root, sterile neutrinos can affect the 
world only through their small vacuum mixing 
with active neutrinos, but this also allows 
several decay modes, one of which is the simple 
beta decay-like mode, where a heavier, nonrelativistic, mostly 
sterile neutrino decays into a light, mostly active neutrino and a photon. 
The rate of this decay scales like five powers of the sterile neutrino rest mass scale, and is proportional to the square of the appropriate active-sterile vacuum mixing angle. Sterile neutrino dark matter candidates with $\sim {\rm keV}$ rest masses produce X-rays via this decay mode, and there are many existing and future X-ray observatories. This is where the best constraints on sterile neutrino dark matter come from~\cite{Joudaki:2012uk}. 
Most of the simplest models for production of a relic sterile neutrino density in a range to be a significant component of the dark matter require vacuum mixing and rest mass parameters that run afoul of the X-ray constraints \cite{Abazajian:2001lr,Abazajian:2001fk,Yuksel:2008fk}. 

However, there are models that would produce the right relic densities for sterile neutrinos, yet can evade all existing cosmological and laboratory bounds. Examples of these include models which rely on matter enhancement \cite{Shi:1999lq}, or models for producing a relic density that are built on Higgs decay \cite{Kusenko:2010qy}. It should be noted that not all effects of sterile neutrinos are necessarily bad.

Sterile neutrinos, mixing in a medium with $\nu_e$'s and $\bar\nu_e$'s, could solve the alpha effect problem in neutrino-heated  $r$-process nucleosynthesis models \cite{McLaughlin:1999fk,Caldwell:2000db,Fetter:2003lr}. This model has the virtue that it can enable active-sterile neutrino medium-enhanced mixing to engineer an 
extreme neutron excess which, in turn, can lead to fission cycling in the r-process. Such cycling may be necessary to explain the observational fact that the nuclear mass 130 peak and 195 peak abundances are comparable, and that is difficult to understand absent some mechanism to drive the equilibrium between the abundances in these mass regions. Fission cycling does this naturally. This is a key point of contact between an outstanding and vexing problem in nuclear physics and astrophysics, {\it i.e.}, the origin of the r-process elements such as 
iodine and uranium, and the speculative physics associated with a possible sterile neutrino sector. As a consequence, nuclear physicists have a vested interest in sterile neutrinos, and not just because these particles could be dark matter candidates. 

Moreover, finding one kind of light sterile neutrino, {\it e.g.}, one with a mass scale $\sim 1\,{\rm eV}$, immediately buttresses the arguments  
for looking for other light sterile species, {\it e.g.}, those with $\sim 1\,{\rm keV}$ mass scales which 
could be a significant component of the dark matter \footnote{One of the authors of the present work has dubbed this argument the \lq\lq Cockroach Principle,\rq\rq\ meaning that if you find one, there are likely to be others. The author of Ref.~\cite{Kusenko:2009lr}, being of Russian origin, deems this the \lq\lq Mushroom Principle,\rq\rq\ because where you find one mushroom there are likely to be others nearby. And you actually {\it want} to find mushrooms, not cockroaches.}. If, for example, nuclear physicists could establish that the r-process cannot operate in supernovae or compact object mergers {\it without} a sterile neutrino, then the interest in sterile neutrino dark matter is heightened across the board. Of course, presently we are nowhere near drawing any such conclusion. All we can say at this point is that with the anticipated advent of Advanced LIGO,  and direct observation of compact object mergers, we will understand more. We can also say that these topics are right in the heart of important frontline issues in nuclear physics.

This is just one example, and certainly not the only one, 
in which sterile neutrino dark matter physics issues overlap with other thrusts in modern nuclear physics. Consider another example, one which overlaps with important physics being studied in relativistic heavy ion collisions and in fundamental lattice QCD calculations. Some models for production of a cosmologically significant sterile neutrino relic density produce that relic density through active-neutrino scattering-induced de-coherence 
in the early universe. 
The production rate in this case 
in the early universe is negligible at 
very high temperatures, where the 
active neutrino scattering rate 
is so high that quantum mechanical 
suppression of active sterile mixing ({\it i.e.}, 
the quantum Zeno effect) is dominant. Likewise, at low temperatures the sterile neutrino production rate is low because the scattering rate is low.

The bulk of sterile neutrino production lies between these scales, in fact right in the QCD epoch, where the temperature scale is $\sim 100\,{\rm MeV}$. At issue is how active neutrinos interact in the dense, hot (high entropy) nuclear matter that comprises the early universe medium at these temperatures. Though the QCD community concentrates, as they should, on studying the bulk properties of this medium, such as the 
baryon number susceptibility, the sterile
 neutrino dark matter models bring up new topics for investigation. For example, what is the active neutrino transport mean free path in this medium? What are the relevant weak interaction degrees of freedom?
There is another way that sterile neutrino dark matter ideas tie together nuclear astrophysics and astronomy. A rapidly developing arena of research is the origin of galaxies and, especially, reconciling ideas on dark matter with this subject, as described above. An unresolved issue there is the chemical (nuclear abundance) evolution of dwarf galaxies and other structures. Understanding this may allow insights into whether the perceived troubles with small-scale structures such as the dwarf spheroidal galaxies stem from lack of understanding of how prosaic processes such as gas physics operate, or whether they come from some key misunderstanding about the nature of the dark matter itself. Examples of the latter include questions of whether dark matter is warm or cold (sterile neutrinos can be either), or self-interacting.

\section{Summary}

Astrophysical observations tell us that we live in a dark-dominated universe, though
the precise mechanisms which give rise to its nature have not yet been determined. 
We have worked under the assumption that particle physics, and particularly 
the physics of the weak scale, might yet explain it. In this context we 
have reviewed the astrophysical observations and simulations and experiments 
which inform us about dark matter. 

Recent results from collider physics, astrophysics, and cosmology 
encourage broader thinking 
in regards to possible dark-matter candidates --- dark-matter need not
be made exclusively of ``WIMPs.'' Facilities dedicated 
to nuclear physics are well-positioned to investigate certain 
non-WIMP models, and we have discussed the models which are probed at
such facilities in some detail. 
In parallel to this, developments in observational cosmology 
permit probes of the relativistic energy density 
at early epochs and thus provide 
new ways to constrain dark-matter models, provided nuclear physics inputs
are sufficiently well-known. 
The emerging confluence of constraints from diverse sources, be they 
accelerator, astrophysical, or cosmological, 
permit searches for dark-matter candidates 
in a greater range of masses 
and interaction strengths than heretofore possible, and we 
 conclude that a bright future exists for the discovery of things dark. 

\section{Acknowledgments}

SG acknowledges partial support from the U.S. Department of Energy under 
contract DE-FG02-96ER40989, and GMF acknowledges partial support
from NSF grant PHY-09-70064 and the 
UC Office of the President. We would like to acknowledge helpful conversations
with K. Abazajian and A. Kusenko, and we thank E. Aprile for providing
the graphic shown in Fig.~\ref{fig:xenon100}. 

%

%

\bibliographystyle{doiplain}
\bibliography{Dark}

\begin{thebibliography}{100}

\bibitem{Hu:2001bc}
Wayne Hu and Scott Dodelson.
\newblock {Cosmic microwave background anisotropies}.
\newblock {\em Ann.Rev.Astron.Astrophys.}, 40:171\unskip--\ignorespaces 216,
  2002, \doi{10.1146/annurev.astro.40.060401.093926},
  \eprint{arXiv}{astro-ph/0110414}.

\bibitem{Eisenstein:2005su}
Daniel~J. Eisenstein et~al., SDSS Collaboration.
\newblock {Detection of the baryon acoustic peak in the large-scale correlation
  function of SDSS luminous red galaxies}.
\newblock {\em Astrophys.J.}, 633:560\unskip--\ignorespaces 574, 2005,
  \doi{10.1086/466512}, \eprint{arXiv}{astro-ph/0501171}.

\bibitem{Faber:1979pp}
S.M. Faber and J.S. Gallagher.
\newblock {Masses and mass-to-light ratios of galaxies}.
\newblock {\em Ann.Rev.Astron.Astrophys.}, 17:135\unskip--\ignorespaces 183,
  1979, \doi{10.1146/annurev.aa.17.090179.001031}.

\bibitem{Rubin:1980zd}
V.C. Rubin, N.~Thonnard, and Jr. Ford, W.K.
\newblock {Rotational properties of 21 SC galaxies with a large range of
  luminosities and radii, from NGC 4605 /R = 4kpc/ to UGC 2885 /R = 122 kpc/}.
\newblock {\em Astrophys.J.}, 238:471, 1980, \doi{10.1086/158003}.

\bibitem{Hinshaw:2012}
G.~Hinshaw et~al., WMAP Collaboration.
\newblock {Nine-Year Wilkinson Microwave Anisotropy Probe (WMAP) Observations:
  Cosmological Parameter Results}.
\newblock 2012, \eprint{arXiv}{1212.5226}.

\bibitem{Riess:1998cb}
Adam~G. Riess et~al., Supernova Search Team.
\newblock {Observational evidence from supernovae for an accelerating universe
  and a cosmological constant}.
\newblock {\em Astron.J.}, 116:1009\unskip--\ignorespaces 1038, 1998,
  \doi{10.1086/300499}, \eprint{arXiv}{astro-ph/9805201}.

\bibitem{Perlmutter:1998np}
S.~Perlmutter et~al., Supernova Cosmology Project.
\newblock {Measurements of Omega and Lambda from 42 high redshift supernovae}.
\newblock {\em Astrophys.J.}, 517:565\unskip--\ignorespaces 586, 1999,
  \doi{10.1086/307221}, \eprint{arXiv}{astro-ph/9812133}.

\bibitem{Clowe:2006eq}
Douglas Clowe, Marusa Bradac, Anthony~H. Gonzalez, Maxim Markevitch, Scott~W.
  Randall, et~al.
\newblock {A direct empirical proof of the existence of dark matter}.
\newblock {\em Astrophys.J.}, 648:L109\unskip--\ignorespaces L113, 2006,
  \doi{10.1086/508162}, \eprint{arXiv}{astro-ph/0608407}.

\bibitem{Bean:2010zq}
Rachel Bean and Matipon Tangmatitham.
\newblock {Current constraints on the cosmic growth history}.
\newblock {\em Phys.Rev.}, D81:083534, 2010, \doi{10.1103/PhysRevD.81.083534},
  \eprint{arXiv}{1002.4197}.

\bibitem{Lombriser:2010mp}
Lucas Lombriser, Anze Slosar, Uros Seljak, and Wayne Hu.
\newblock {Constraints on f(R) gravity from probing the large-scale structure}.
\newblock {\em Phys.Rev.}, D85:124038, 2012, \doi{10.1103/PhysRevD.85.124038},
  \eprint{arXiv}{1003.3009}.

\bibitem{Weinberg:2012es}
David~H. Weinberg, Michael~J. Mortonson, Daniel~J. Eisenstein, Christopher
  Hirata, Adam~G. Riess, et~al.
\newblock {Observational Probes of Cosmic Acceleration}.
\newblock 2012, \eprint{arXiv}{1201.2434}.

\bibitem{Bertone:2004pz}
Gianfranco Bertone, Dan Hooper, and Joseph Silk.
\newblock {Particle dark matter: Evidence, candidates and constraints}.
\newblock {\em Phys.Rept.}, 405:279\unskip--\ignorespaces 390, 2005,
  \doi{10.1016/j.physrep.2004.08.031}, \eprint{arXiv}{hep-ph/0404175}.

\bibitem{Gaitskell:2004gd}
R.J. Gaitskell.
\newblock {Direct detection of dark matter}.
\newblock {\em Ann.Rev.Nucl.Part.Sci.}, 54:315\unskip--\ignorespaces 359, 2004,
  \doi{10.1146/annurev.nucl.54.070103.181244}.

\bibitem{Jaeckel:2010ni}
Joerg Jaeckel and Andreas Ringwald.
\newblock {The Low-Energy Frontier of Particle Physics}.
\newblock {\em Ann.Rev.Nucl.Part.Sci.}, 60:405\unskip--\ignorespaces 437, 2010,
  \doi{10.1146/annurev.nucl.012809.104433}, \eprint{arXiv}{1002.0329}.

\bibitem{Feng:2010gw}
Jonathan~L. Feng.
\newblock {Dark Matter Candidates from Particle Physics and Methods of
  Detection}.
\newblock {\em Ann.Rev.Astron.Astrophys.}, 48:495\unskip--\ignorespaces 545,
  2010, \doi{10.1146/annurev-astro-082708-101659}, \eprint{arXiv}{1003.0904}.

\bibitem{Feng:2010tg}
Jonathan~L. Feng.
\newblock {Non-WIMP Candidates}.
\newblock 2010, \eprint{arXiv}{1002.3828}.

\bibitem{atlas:2012gk}
Georges Aad et~al., ATLAS Collaboration.
\newblock {Observation of a new particle in the search for the Standard Model
  Higgs boson with the ATLAS detector at the LHC}.
\newblock 2012, \eprint{arXiv}{1207.7214}.

\bibitem{cms:2012gu}
Serguei Chatrchyan et~al., CMS Collaboration.
\newblock {Observation of a new boson at a mass of 125 GeV with the CMS
  experiment at the LHC}.
\newblock {\em Phys.Lett.B}, 2012, \eprint{arXiv}{1207.7235}.

\bibitem{Feng:2012rn}
Jonathan~L. Feng, Ze'ev Surujon, and Hai-Bo Yu.
\newblock {Confluence of Constraints in Gauge Mediation: The 125 GeV Higgs
  Boson and Goldilocks Cosmology}.
\newblock 2012, \eprint{arXiv}{1205.6480}.

\bibitem{Murayama:2012jh}
Hitoshi Murayama, Yasunori Nomura, Satoshi Shirai, and Kohsaku Tobioka.
\newblock {Compact Supersymmetry}.
\newblock {\em Phys.Rev.}, D86:115014, 2012, \doi{10.1103/PhysRevD.86.115014},
  \eprint{arXiv}{1206.4993}.

\bibitem{Feng:2008ya}
Jonathan~L. Feng and Jason Kumar.
\newblock {The WIMPless Miracle: Dark-Matter Particles without Weak-Scale
  Masses or Weak Interactions}.
\newblock {\em Phys.Rev.Lett.}, 101:231301, 2008,
  \doi{10.1103/PhysRevLett.101.231301}, \eprint{arXiv}{0803.4196}.

\bibitem{Planck:2006aa}
Planck Collaboration.
\newblock {The Scientific programme of planck}.
\newblock 2006, \eprint{arXiv}{astro-ph/0604069}.

\bibitem{Ade:2011ah}
P.A.R. Ade et~al., Planck Collaboration.
\newblock {Planck Early Results. I. The Planck mission}.
\newblock {\em Astron.Astrophys.}, 536:16464, 2011,
  \doi{10.1051/0004-6361/201116464}, \eprint{arXiv}{1101.2022}.

\bibitem{Steigman:2007xt}
Gary Steigman.
\newblock {Primordial Nucleosynthesis in the Precision Cosmology Era}.
\newblock {\em Ann.Rev.Nucl.Part.Sci.}, 57:463\unskip--\ignorespaces 491, 2007,
  \doi{10.1146/annurev.nucl.56.080805.140437}, \eprint{arXiv}{0712.1100}.

\bibitem{Steigman:2012ve}
Gary Steigman.
\newblock {Neutrinos And Big Bang Nucleosynthesis}.
\newblock 2012, \eprint{arXiv}{1208.0032}.

\bibitem{Abazajian:2011dt}
K.N. Abazajian, E.~Calabrese, A.~Cooray, F.~De~Bernardis, S.~Dodelson, et~al.
\newblock {Cosmological and Astrophysical Neutrino Mass Measurements}.
\newblock {\em Astropart.Phys.}, 35:177\unskip--\ignorespaces 184, 2011,
  \doi{10.1016/j.astropartphys.2011.07.002}, \eprint{arXiv}{1103.5083}.

\bibitem{Fuller:2011qy}
George~M. Fuller, Chad~T. Kishimoto, and Alexander Kusenko.
\newblock {Heavy sterile neutrinos, entropy and relativistic energy production,
  and the relic neutrino background}.
\newblock 2011, \eprint{arXiv}{1110.6479}.

\bibitem{Harnik:2012ni}
Roni Harnik, Joachim Kopp, and Pedro~A.N. Machado.
\newblock {Exploring nu Signals in Dark Matter Detectors}.
\newblock {\em JCAP}, 1207:026, 2012, \doi{10.1088/1475-7516/2012/07/026},
  \eprint{arXiv}{1202.6073}.

\bibitem{Boyd:2010kj}
Richard~N. Boyd, Carl~R. Brune, George~M. Fuller, and Christel~J. Smith.
\newblock {New Nuclear Physics for Big Bang Nucleosynthesis}.
\newblock {\em Phys.Rev.}, D82:105005, 2010, \doi{10.1103/PhysRevD.82.105005},
  \eprint{arXiv}{1008.0848}.

\bibitem{greene}
Geoffrey~L. Greene and Fred~E. Wietfeldt.
\newblock {Colloquium: The neutron lifetime}.
\newblock {\em Rev.Mod.Phys.}, 83:1173, 2011.

\bibitem{PDG2012}
J.~Beringer et~al., Particle Data Group.
\newblock {Review of Particle Physics}.
\newblock {\em Phys.Rev.D}, 86:010001, 2012.

\bibitem{Nollett:2011aa}
Kenneth~M. Nollett and Gilbert~P. Holder.
\newblock {An analysis of constraints on relativistic species from primordial
  nucleosynthesis and the cosmic microwave background}.
\newblock 2011, \eprint{arXiv}{1112.2683}.

\bibitem{Smith:2006uw}
Christel~J. Smith, George~M. Fuller, Chad~T. Kishimoto, and Kevork~N.
  Abazajian.
\newblock {Light Element Signatures of Sterile Neutrinos and Cosmological
  Lepton Numbers}.
\newblock {\em Phys.Rev.}, D74:085008, 2006, \doi{10.1103/PhysRevD.74.085008},
  \eprint{arXiv}{astro-ph/0608377}.

\bibitem{White:1984yj}
Simon~D.M. White, C.S. Frenk, and M.~Davis.
\newblock {Clustering in a Neutrino Dominated Universe}.
\newblock {\em Astrophys.J.}, 274:L1\unskip--\ignorespaces L5, 1983.

\bibitem{Sikivie:2006ni}
Pierre Sikivie.
\newblock {Axion Cosmology}.
\newblock {\em Lect.Notes Phys.}, 741:19\unskip--\ignorespaces 50, 2008,
  \doi{10.1007/978-3-540-73518-2${\underline{\,}}$2},
  \eprint{arXiv}{astro-ph/0610440}.

\bibitem{Blumenthal:1984bp}
George~R. Blumenthal, S.M. Faber, Joel~R. Primack, and Martin~J. Rees.
\newblock {Formation of Galaxies and Large Scale Structure with Cold Dark
  Matter}.
\newblock {\em Nature}, 311:517\unskip--\ignorespaces 525, 1984,
  \doi{10.1038/311517a0}.

\bibitem{Press:1973iz}
William~H. Press and Paul Schechter.
\newblock {Formation of galaxies and clusters of galaxies by selfsimilar
  gravitational condensation}.
\newblock {\em Astrophys.J.}, 187:425\unskip--\ignorespaces 438, 1974,
  \doi{10.1086/152650}.

\bibitem{White:1977jf}
Simon~D.M. White and M.J. Rees.
\newblock {Core condensation in heavy halos: A Two stage theory for galaxy
  formation and clusters}.
\newblock {\em Mon.Not.Roy.Astron.Soc.}, 183:341\unskip--\ignorespaces 358,
  1978.

\bibitem{Davis:1985rj}
Marc Davis, George Efstathiou, Carlos~S. Frenk, and Simon~D.M. White.
\newblock {The Evolution of Large Scale Structure in a Universe Dominated by
  Cold Dark Matter}.
\newblock {\em Astrophys.J.}, 292:371\unskip--\ignorespaces 394, 1985,
  \doi{10.1086/163168}.

\bibitem{zeldovichp}
Ia.B Zeldovich.
\newblock {The theory of the large scale structure of the universe}.
\newblock {\em The large scale structure of the universe; Proceedings of the
  Symposium, Tallin, Estonian SSR, September 12-16, 1977.}, 1978.

\bibitem{Bond:1980ha}
J.R. Bond, G.~Efstathiou, and J.~Silk.
\newblock {Massive Neutrinos and the Large Scale Structure of the Universe}.
\newblock {\em Phys.Rev.Lett.}, 45:1980\unskip--\ignorespaces 1984, 1980,
  \doi{10.1103/PhysRevLett.45.1980}.

\bibitem{Bond:1988ez}
J.R. Bond, A.S. Szalay, J.~Centrella, and J.R. Wilson.
\newblock {DARK MATTER AND SHOCKED PANCAKES}.
\newblock {\em Formation and Evolution of Galaxies and Large Structures in the
  Universe}, 1984.

\bibitem{Kauffmann:1995gi}
Guinevere Kauffmann.
\newblock {Disc galaxies at z=0 and at high redshift: an explanation of the
  observed evolution of damped Lyalpha absorption systems}.
\newblock {\em Mon.Not.Roy.Astron.Soc.}, 281:475, 1996,
  \eprint{arXiv}{astro-ph/9512123}.

\bibitem{Prochaska:1997xi}
Jason~X. Prochaska and Arthur~M. Wolfe.
\newblock {On the Kinematics of the damped Lyman-alpha protogalaxies}.
\newblock {\em Astrophys.J.}, 487:73, 1997, \doi{10.1086/304591},
  \eprint{arXiv}{astro-ph/9704169}.

\bibitem{Tremaine:1979we}
S.~Tremaine and J.E. Gunn.
\newblock {Dynamical Role of Light Neutral Leptons in Cosmology}.
\newblock {\em Phys.Rev.Lett.}, 42:407\unskip--\ignorespaces 410, 1979,
  \doi{10.1103/PhysRevLett.42.407}.

\bibitem{Kauffmann:1993gv}
G~Kauffmann, Simon~D.M. White, and B.~Guiderdoni.
\newblock {The Formation and Evolution of Galaxies Within Merging Dark Matter
  Haloes}.
\newblock {\em Mon.Not.Roy.Astron.Soc.}, 264:201, 1993.

\bibitem{Klypin:1999uc}
Anatoly~A. Klypin, Andrey~V. Kravtsov, Octavio Valenzuela, and Francisco Prada.
\newblock {Where are the missing Galactic satellites?}
\newblock {\em Astrophys.J.}, 522:82\unskip--\ignorespaces 92, 1999,
  \doi{10.1086/307643}, \eprint{arXiv}{astro-ph/9901240}.

\bibitem{Moore:1999nt}
B.~Moore, S.~Ghigna, F.~Governato, G.~Lake, Thomas~R. Quinn, et~al.
\newblock {Dark matter substructure within galactic halos}.
\newblock {\em Astrophys.J.}, 524:L19\unskip--\ignorespaces L22, 1999,
  \doi{10.1086/312287}, \eprint{arXiv}{astro-ph/9907411}.

\bibitem{Bullock:2010uy}
James~S. Bullock.
\newblock {Notes on the Missing Satellites Problem}.
\newblock 2010, \eprint{arXiv}{1009.4505}.

\bibitem{Bode:2000gq}
Paul Bode, Jeremiah~P. Ostriker, and Neil Turok.
\newblock {Halo formation in warm dark matter models}.
\newblock {\em Astrophys.J.}, 556:93\unskip--\ignorespaces 107, 2001,
  \doi{10.1086/321541}, \eprint{arXiv}{astro-ph/0010389}.

\bibitem{Moore:1999gc}
Ben Moore, Thomas~R. Quinn, Fabio Governato, Joachim Stadel, and George Lake.
\newblock {Cold collapse and the core catastrophe}.
\newblock {\em Mon.Not.Roy.Astron.Soc.}, 310:1147\unskip--\ignorespaces 1152,
  1999, \doi{10.1046/j.1365-8711.1999.03039.x},
  \eprint{arXiv}{astro-ph/9903164}.

\bibitem{AvilaReese:2000hg}
Vladimir Avila-Reese, Pefro Colin, Octavio Valenzuela, Elena D'Onghia, and
  Claudio Firmani.
\newblock {Formation and structure of halos in a warm dark matter cosmology}.
\newblock {\em Astrophys.J.}, 559:516\unskip--\ignorespaces 530, 2001,
  \doi{10.1086/322411}, \eprint{arXiv}{astro-ph/0010525}.

\bibitem{Viel:2005qj}
Matteo Viel, Julien Lesgourgues, Martin~G. Haehnelt, Sabino Matarrese, and
  Antonio Riotto.
\newblock {Constraining warm dark matter candidates including sterile neutrinos
  and light gravitinos with WMAP and the Lyman-alpha forest}.
\newblock {\em Phys.Rev.}, D71:063534, 2005, \doi{10.1103/PhysRevD.71.063534},
  \eprint{arXiv}{astro-ph/0501562}.

\bibitem{Seljak:2006qw}
Uros Seljak, Alexey Makarov, Patrick McDonald, and Hy~Trac.
\newblock {Can sterile neutrinos be the dark matter?}
\newblock {\em Phys.Rev.Lett.}, 97:191303, 2006,
  \doi{10.1103/PhysRevLett.97.191303}, \eprint{arXiv}{astro-ph/0602430}.

\bibitem{Viel:2006kd}
Matteo Viel, Julien Lesgourgues, Martin~G. Haehnelt, Sabino Matarrese, and
  Antonio Riotto.
\newblock {Can sterile neutrinos be ruled out as warm dark matter candidates?}
\newblock {\em Phys.Rev.Lett.}, 97:071301, 2006,
  \doi{10.1103/PhysRevLett.97.071301}, \eprint{arXiv}{astro-ph/0605706}.

\bibitem{Viel:2007mv}
Matteo Viel, George~D. Becker, James~S. Bolton, Martin~G. Haehnelt, Michael
  Rauch, et~al.
\newblock {How cold is cold dark matter? Small scales constraints from the flux
  power spectrum of the high-redshift Lyman-alpha forest}.
\newblock {\em Phys.Rev.Lett.}, 100:041304, 2008,
  \doi{10.1103/PhysRevLett.100.041304}, \eprint{arXiv}{0709.0131}.

\bibitem{Boyarsky:2008xj}
Alexey Boyarsky, Julien Lesgourgues, Oleg Ruchayskiy, and Matteo Viel.
\newblock {Lyman-alpha constraints on warm and on warm-plus-cold dark matter
  models}.
\newblock {\em JCAP}, 0905:012, 2009, \doi{10.1088/1475-7516/2009/05/012},
  \eprint{arXiv}{0812.0010}.

\bibitem{Petraki:2007gq}
Kalliopi Petraki and Alexander Kusenko.
\newblock {Dark-matter sterile neutrinos in models with a gauge singlet in the
  Higgs sector}.
\newblock {\em Phys.Rev.}, D77:065014, 2008, \doi{10.1103/PhysRevD.77.065014},
  \eprint{arXiv}{0711.4646}.

\bibitem{Hut:1977zn}
P.~Hut.
\newblock {Limits on Masses and Number of Neutral Weakly Interacting
  Particles}.
\newblock {\em Phys.Lett.}, B69:85, 1977, \doi{10.1016/0370-2693(77)90139-3}.

\bibitem{Lee:1977ua}
Benjamin~W. Lee and Steven Weinberg.
\newblock {Cosmological Lower Bound on Heavy Neutrino Masses}.
\newblock {\em Phys.Rev.Lett.}, 39:165\unskip--\ignorespaces 168, 1977,
  \doi{10.1103/PhysRevLett.39.165}.

\bibitem{Vysotsky:1977pe}
M.I. Vysotsky, A.D. Dolgov, and Ya.B. Zeldovich.
\newblock {Cosmological Restriction on Neutral Lepton Masses}.
\newblock {\em JETP Lett.}, 26:188\unskip--\ignorespaces 190, 1977.

\bibitem{Feng:2008mu}
Jonathan~L. Feng, Huitzu Tu, and Hai-Bo Yu.
\newblock {Thermal Relics in Hidden Sectors}.
\newblock {\em JCAP}, 0810:043, 2008, \doi{10.1088/1475-7516/2008/10/043},
  \eprint{arXiv}{0808.2318}.

\bibitem{Adriani:2008zr}
Oscar Adriani et~al., PAMELA Collaboration.
\newblock {An anomalous positron abundance in cosmic rays with energies 1.5-100
  GeV}.
\newblock {\em Nature}, 458:607\unskip--\ignorespaces 609, 2009,
  \doi{10.1038/nature07942}, \eprint{arXiv}{0810.4995}.

\bibitem{Adriani:2010ib}
O.~Adriani, G.C. Barbarino, G.A. Bazilevskaya, R.~Bellotti, M.~Boezio, et~al.
\newblock {A statistical procedure for the identification of positrons in the
  PAMELA experiment}.
\newblock {\em Astropart.Phys.}, 34:1\unskip--\ignorespaces 11, 2010,
  \doi{10.1016/j.astropartphys.2010.04.007}, \eprint{arXiv}{1001.3522}.

\bibitem{FermiLAT:2011ab}
M.~Ackermann et~al., Fermi LAT Collaboration.
\newblock {Measurement of separate cosmic-ray electron and positron spectra
  with the Fermi Large Area Telescope}.
\newblock {\em Phys.Rev.Lett.}, 108:011103, 2012,
  \doi{10.1103/PhysRevLett.108.011103}, \eprint{arXiv}{1109.0521}.

\bibitem{Arvanitaki:2009yb}
Asimina Arvanitaki, Savas Dimopoulos, Sergei Dubovsky, Peter~W. Graham, Roni
  Harnik, et~al.
\newblock {Decaying Dark Matter as a Probe of Unification and TeV
  Spectroscopy}.
\newblock {\em Phys.Rev.}, D80:055011, 2009, \doi{10.1103/PhysRevD.80.055011},
  \eprint{arXiv}{0904.2789}.

\bibitem{Spergel:1999mh}
David~N. Spergel and Paul~J. Steinhardt.
\newblock {Observational evidence for selfinteracting cold dark matter}.
\newblock {\em Phys.Rev.Lett.}, 84:3760\unskip--\ignorespaces 3763, 2000,
  \doi{10.1103/PhysRevLett.84.3760}, \eprint{arXiv}{astro-ph/9909386}.

\bibitem{Dimopoulos:1989hk}
Savas Dimopoulos, David Eichler, Rahim Esmailzadeh, and Glenn~D. Starkman.
\newblock {GETTING A CHARGE OUT OF DARK MATTER}.
\newblock {\em Phys.Rev.}, D41:2388, 1990, \doi{10.1103/PhysRevD.41.2388}.

\bibitem{Gould:1989gw}
Andrew Gould, Bruce~T. Draine, Roger~W. Romani, and Shmuel Nussinov.
\newblock {Neutron Stars: Graveyard of Charged Dark Matter}.
\newblock {\em Phys.Lett.}, B238:337, 1990, \doi{10.1016/0370-2693(90)91745-W}.

\bibitem{Gradwohl:1992ue}
Ben-Ami Gradwohl and Joshua~A. Frieman.
\newblock {Dark matter, long range forces, and large scale structure}.
\newblock {\em Astrophys.J.}, 398:407\unskip--\ignorespaces 424, 1992,
  \doi{10.1086/171865}.

\bibitem{Rocha:2012jg}
Miguel Rocha, Annika~H.G. Peter, James~S. Bullock, Manoj Kaplinghat, Shea
  Garrison-Kimmel, et~al.
\newblock {Cosmological Simulations with Self-Interacting Dark Matter I:
  Constant Density Cores and Substructure}.
\newblock 2012, \eprint{arXiv}{1208.3025}.

\bibitem{Peter:2012jh}
Annika~H.G. Peter, Miguel Rocha, James~S. Bullock, and Manoj Kaplinghat.
\newblock {Cosmological Simulations with Self-Interacting Dark Matter II: Halo
  Shapes vs. Observations}.
\newblock 2012, \eprint{arXiv}{1208.3026}.

\bibitem{Kirkman:2003uv}
David Kirkman, David Tytler, Nao Suzuki, John~M. O'Meara, and Dan Lubin.
\newblock {The Cosmological baryon density from the deuterium to hydrogen ratio
  towards QSO absorption systems: D/H towards Q1243+3047}.
\newblock {\em Astrophys.J.Suppl.}, 149:1, 2003, \doi{10.1086/378152},
  \eprint{arXiv}{astro-ph/0302006}.

\bibitem{Mangano:2005cc}
Gianpiero Mangano, Gennaro Miele, Sergio Pastor, Teguayco Pinto, Ofelia
  Pisanti, et~al.
\newblock {Relic neutrino decoupling including flavor oscillations}.
\newblock {\em Nucl.Phys.}, B729:221\unskip--\ignorespaces 234, 2005,
  \doi{10.1016/j.nuclphysb.2005.09.041}, \eprint{arXiv}{hep-ph/0506164}.

\bibitem{Hou:2012xq}
Z.~Hou, C.L. Reichardt, K.T. Story, B.~Follin, R.~Keisler, et~al.
\newblock {Constraints on Cosmology from the Cosmic Microwave Background Power
  Spectrum of the 2500-square degree SPT-SZ Survey}.
\newblock 2012, \eprint{arXiv}{1212.6267}.

\bibitem{Sievers:2013wk}
Jonathan~L. Sievers, Renee~A. Hlozek, Michael~R. Nolta, Viviana Acquaviva,
  Graeme~E. Addison, et~al.
\newblock {The Atacama Cosmology Telescope: Cosmological parameters from three
  seasons of data}.
\newblock 2013, \eprint{arXiv}{1301.0824}.

\bibitem{Alcock:2000ph}
C.~Alcock et~al., MACHO Collaboration.
\newblock {The MACHO project: Microlensing results from 5.7 years of LMC
  observations}.
\newblock {\em Astrophys.J.}, 542:281\unskip--\ignorespaces 307, 2000,
  \doi{10.1086/309512}, \eprint{arXiv}{astro-ph/0001272}.

\bibitem{Alcock:1998fx}
C.~Alcock et~al., MACHO Collaboration, EROS Collaboration.
\newblock {EROS and MACHO combined limits on planetary mass dark matter in the
  galactic halo}.
\newblock {\em Astrophys.J.Lett.}, 1998, \eprint{arXiv}{astro-ph/9803082}.

\bibitem{Tisserand:2006zx}
P.~Tisserand et~al., EROS-2 Collaboration.
\newblock {Limits on the Macho Content of the Galactic Halo from the EROS-2
  Survey of the Magellanic Clouds}.
\newblock {\em Astron.Astrophys.}, 469:387\unskip--\ignorespaces 404, 2007,
  \doi{10.1051/0004-6361:20066017}, \eprint{arXiv}{astro-ph/0607207}.

\bibitem{Wyrzykowski:2011tr}
L.~Wyrzykowski, J.~Skowron, S.~Kozlowski, A.~Udalski, M.K. Szymanski, et~al.
\newblock {The OGLE View of Microlensing towards the Magellanic Clouds. IV.
  OGLE-III SMC Data and Final Conclusions on MACHOs}.
\newblock {\em Mon.Not.Roy.Astron.Soc.}, 416:2949\unskip--\ignorespaces 2961,
  2011, \eprint{arXiv}{1106.2925}.

\bibitem{Cieplak:2012dp}
Agnieszka~M. Cieplak and Kim Griest.
\newblock {Improved Theoretical Predictions of Microlensing Rates for the
  Detection of Primordial Black Hole Dark Matter}.
\newblock 2012, \eprint{arXiv}{1210.7729}.

\bibitem{dmsag2007}
E.-K. Park.
\newblock {contribution to DMSAG report, July 18, 2007.}
\newblock {\em http://science.energy.gov/hep/hepap/reports/}, 2007.

\bibitem{Drukier:1983gj}
A.~Drukier and Leo Stodolsky.
\newblock {Principles and Applications of a Neutral Current Detector for
  Neutrino Physics and Astronomy}.
\newblock {\em Phys.Rev.}, D30:2295, 1984, \doi{10.1103/PhysRevD.30.2295}.

\bibitem{Goodman:1984dc}
Mark~W. Goodman and Edward Witten.
\newblock {Detectability of Certain Dark Matter Candidates}.
\newblock {\em Phys.Rev.}, D31:3059, 1985, \doi{10.1103/PhysRevD.31.3059}.

\bibitem{Sikivie:1985yu}
P.~Sikivie.
\newblock {DETECTION RATES FOR 'INVISIBLE' AXION SEARCHES}.
\newblock {\em Phys.Rev.}, D32:2988, 1985, \doi{10.1103/PhysRevD.36.974,
  10.1103/PhysRevD.32.2988}.

\bibitem{Asztalos:2009yp}
S.J. Asztalos et~al., ADMX Collaboration.
\newblock {A SQUID-based microwave cavity search for dark-matter axions}.
\newblock {\em Phys.Rev.Lett.}, 104:041301, 2010,
  \doi{10.1103/PhysRevLett.104.041301}, \eprint{arXiv}{0910.5914}.

\bibitem{Gardner:2006za}
Susan Gardner.
\newblock {Observing Dark Matter via the Gyromagnetic Faraday Effect}.
\newblock {\em Phys.Rev.Lett.}, 100:041303, 2008,
  \doi{10.1103/PhysRevLett.100.041303}, \eprint{arXiv}{astro-ph/0611684}.

\bibitem{Gardner:2008yn}
Susan Gardner.
\newblock {Shedding Light on Dark Matter: A Faraday Rotation Experiment to
  Limit a Dark Magnetic Moment}.
\newblock {\em Phys.Rev.}, D79:055007, 2009, \doi{10.1103/PhysRevD.79.055007},
  \eprint{arXiv}{0811.0967}.

\bibitem{Pospelov:2000bq}
Maxim Pospelov and Tonnis ter Veldhuis.
\newblock {Direct and indirect limits on the electromagnetic form-factors of
  WIMPs}.
\newblock {\em Phys.Lett.}, B480:181\unskip--\ignorespaces 186, 2000,
  \doi{10.1016/S0370-2693(00)00358-0}, \eprint{arXiv}{hep-ph/0003010}.

\bibitem{Fan:2010gt}
JiJi Fan, Matthew Reece, and Lian-Tao Wang.
\newblock {Non-relativistic effective theory of dark matter direct detection}.
\newblock {\em JCAP}, 1011:042, 2010, \doi{10.1088/1475-7516/2010/11/042},
  \eprint{arXiv}{1008.1591}.

\bibitem{Fitzpatrick:2012ix}
A.~Liam Fitzpatrick, Wick Haxton, Emanuel Katz, Nicholas Lubbers, and Yiming
  Xu.
\newblock {The Effective Field Theory of Dark Matter Direct Detection}.
\newblock {\em JCAP}, 1302:004, 2013, \doi{10.1088/1475-7516/2013/02/004},
  \eprint{arXiv}{1203.3542}.

\bibitem{Fitzpatrick:2012ib}
A.~Liam Fitzpatrick, Wick Haxton, Emanuel Katz, Nicholas Lubbers, and Yiming
  Xu.
\newblock {Model Independent Direct Detection Analyses}.
\newblock 2012, \eprint{arXiv}{1211.2818}.

\bibitem{Kelso:2011gd}
Chris Kelso, Dan Hooper, and Matthew~R. Buckley.
\newblock {Toward A Consistent Picture For CRESST, CoGeNT and DAMA}.
\newblock {\em Phys.Rev.}, D85:043515, 2012, \doi{10.1103/PhysRevD.85.043515},
  \eprint{arXiv}{1110.5338}.

\bibitem{Khriplovich:2006sq}
Iosif~B. Khriplovich and E.V. Pitjeva.
\newblock {Upper limits on density of dark matter in solar system}.
\newblock {\em Int.J.Mod.Phys.}, D15:615\unskip--\ignorespaces 618, 2006,
  \doi{10.1142/S0218271806008462, 10.1142/9789812834300$\underline{\,}$0053},
  \eprint{arXiv}{astro-ph/0601422}.

\bibitem{Frere:2007pi}
J.-M. Frere, Fu-Sin Ling, and G.~Vertongen.
\newblock {Bound on the Dark Matter Density in the Solar System from Planetary
  Motions}.
\newblock {\em Phys.Rev.}, D77:083005, 2008, \doi{10.1103/PhysRevD.77.083005},
  \eprint{arXiv}{astro-ph/0701542}.

\bibitem{Peter:2009ak}
Annika~H.G. Peter.
\newblock {Getting the astrophysics and particle physics of dark matter out of
  next-generation direct detection experiments}.
\newblock {\em Phys.Rev.}, D81:087301, 2010, \doi{10.1103/PhysRevD.81.087301},
  \eprint{arXiv}{0910.4765}.

\bibitem{Fox:2010bu}
Patrick~J. Fox, Graham~D. Kribs, and Tim~M.P. Tait.
\newblock {Interpreting Dark Matter Direct Detection Independently of the Local
  Velocity and Density Distribution}.
\newblock {\em Phys.Rev.}, D83:034007, 2011, \doi{10.1103/PhysRevD.83.034007},
  \eprint{arXiv}{1011.1910}.

\bibitem{Fox:2010bz}
Patrick~J. Fox, Jia Liu, and Neal Weiner.
\newblock {Integrating Out Astrophysical Uncertainties}.
\newblock {\em Phys.Rev.}, D83:103514, 2011, \doi{10.1103/PhysRevD.83.103514},
  \eprint{arXiv}{1011.1915}.

\bibitem{Peter:2011eu}
Annika~H.G. Peter.
\newblock {WIMP astronomy and particle physics with liquid-noble and cryogenic
  direct-detection experiments}.
\newblock {\em Phys.Rev.}, D83:125029, 2011, \doi{10.1103/PhysRevD.83.125029},
  \eprint{arXiv}{1103.5145}.

\bibitem{Friedland:2012fa}
Alexander Friedland and Ian~M. Shoemaker.
\newblock {Integrating In Dark Matter Astrophysics at Direct Detection
  Experiments}.
\newblock 2012, \eprint{arXiv}{1212.4139}.

\bibitem{Lewin:1995rx}
J.D. Lewin and P.F. Smith.
\newblock {Review of mathematics, numerical factors, and corrections for dark
  matter experiments based on elastic nuclear recoil}.
\newblock {\em Astropart.Phys.}, 6:87\unskip--\ignorespaces 112, 1996,
  \doi{10.1016/S0927-6505(96)00047-3}.

\bibitem{Aprile:2012nq}
E.~Aprile et~al., XENON100 Collaboration.
\newblock {Dark Matter Results from 225 Live Days of XENON100 Data}.
\newblock {\em Phys.Rev.Lett.}, 109:181301, 2012,
  \doi{10.1103/PhysRevLett.109.181301}, \eprint{arXiv}{1207.5988}.

\bibitem{Green:2002ht}
Anne~M. Green.
\newblock {Effect of halo modeling on WIMP exclusion limits}.
\newblock {\em Phys.Rev.}, D66:083003, 2002, \doi{10.1103/PhysRevD.66.083003},
  \eprint{arXiv}{astro-ph/0207366}.

\bibitem{Diemand:2008in}
J.~Diemand, M.~Kuhlen, P.~Madau, M.~Zemp, B.~Moore, et~al.
\newblock {Clumps and streams in the local dark matter distribution}.
\newblock {\em Nature}, 454:735\unskip--\ignorespaces 738, 2008,
  \doi{10.1038/nature07153}, \eprint{arXiv}{0805.1244}.

\bibitem{Springel:2008cc}
Volker Springel, Jie Wang, Mark Vogelsberger, Aaron Ludlow, Adrian Jenkins,
  et~al.
\newblock {The Aquarius Project: the subhalos of galactic halos}.
\newblock {\em Mon.Not.Roy.Astron.Soc.}, 391:1685\unskip--\ignorespaces 1711,
  2008, \doi{10.1111/j.1365-2966.2008.14066.x}, \eprint{arXiv}{0809.0898}.

\bibitem{Bruch:2008rx}
T.~Bruch, J.~Read, L.~Baudis, and G.~Lake.
\newblock {Detecting the Milky Way's Dark Disk}.
\newblock {\em Astrophys.J.}, 696:920\unskip--\ignorespaces 923, 2009,
  \doi{10.1088/0004-637X/696/1/920}, \eprint{arXiv}{0804.2896}.

\bibitem{Purcell:2009yp}
Chris~W. Purcell, James~S. Bullock, and Manoj Kaplinghat.
\newblock {The Dark Disk of the Milky Way}.
\newblock {\em Astrophys.J.}, 703:2275\unskip--\ignorespaces 2284, 2009,
  \doi{10.1088/0004-637X/703/2/2275}, \eprint{arXiv}{0906.5348}.

\bibitem{Green:2010gw}
Anne~M. Green.
\newblock {Dependence of direct detection signals on the WIMP velocity
  distribution}.
\newblock {\em JCAP}, 1010:034, 2010, \doi{10.1088/1475-7516/2010/10/034},
  \eprint{arXiv}{1009.0916}.

\bibitem{Lisanti:2011as}
Mariangela Lisanti and David~N. Spergel.
\newblock {Dark Matter Debris Flows in the Milky Way}.
\newblock {\em Phys.Dark Univ.}, 1:155\unskip--\ignorespaces 161, 2012,
  \doi{10.1016/j.dark.2012.10.007}, \eprint{arXiv}{1105.4166}.

\bibitem{Kuhlen:2012fz}
Michael Kuhlen, Mariangela Lisanti, and David~N. Spergel.
\newblock {Direct Detection of Dark Matter Debris Flows}.
\newblock {\em Phys.Rev.}, D86:063505, 2012, \doi{10.1103/PhysRevD.86.063505},
  \eprint{arXiv}{1202.0007}.

\bibitem{Stiff:2001dq}
David Stiff, Lawrence~M. Widrow, and Joshua Frieman.
\newblock {Signatures of hierarchical clustering in dark matter detection
  experiments}.
\newblock {\em Phys.Rev.}, D64:083516, 2001, \doi{10.1103/PhysRevD.64.083516},
  \eprint{arXiv}{astro-ph/0106048}.

\bibitem{Freese:2012xd}
Katherine Freese, Mariangela Lisanti, and Christopher Savage.
\newblock {Annual Modulation of Dark Matter: A Review}.
\newblock 2012, \eprint{arXiv}{1209.3339}.

\bibitem{Purcell:2011nf}
Chris~W. Purcell, James~S. Bullock, Erik Tollerud, Miguel Rocha, and Sukanya
  Chakrabarti.
\newblock {The Sagittarius impact as an architect of spirality and outer rings
  in the Milky Way}.
\newblock {\em Nature}, 477:301\unskip--\ignorespaces 303, 2011,
  \doi{10.1038/nature10417}, \eprint{arXiv}{1109.2918}.

\bibitem{Purcell:2012sh}
Chris~W. Purcell, Andrew~R. Zentner, and Mei-Yu Wang.
\newblock {Dark Matter Direct Search Rates in Simulations of the Milky Way and
  Sagittarius Stream}.
\newblock {\em JCAP}, 1208:027, 2012, \doi{10.1088/1475-7516/2012/08/027},
  \eprint{arXiv}{1203.6617}.

\bibitem{Widrow:2012wu}
Lawrence~M. Widrow, Susan Gardner, Brian Yanny, Scott Dodelson, and Hsin-Yu
  Chen.
\newblock {Galactoseismology: Discovery of Vertical Waves in the Galactic
  Disk}.
\newblock {\em Astrophys.J.}, 750:L41, 2012, \doi{10.1088/2041-8205/750/2/L41},
  \eprint{arXiv}{1203.6861}.

\bibitem{Gomez:2012rd}
Facundo~A. Gomez, Ivan Minchev, Brian~W. O'Shea, Timothy~C. Beers, James~S.
  Bullock, et~al.
\newblock {Vertical density waves in the Milky Way disc induced by the
  Sagittarius Dwarf Galaxy}.
\newblock 2012, \eprint{arXiv}{1207.3083}.

\bibitem{Bergstrom:1997fj}
Lars Bergstrom, Piero Ullio, and James~H. Buckley.
\newblock {Observability of gamma-rays from dark matter neutralino
  annihilations in the Milky Way halo}.
\newblock {\em Astropart.Phys.}, 9:137\unskip--\ignorespaces 162, 1998,
  \doi{10.1016/S0927-6505(98)00015-2}, \eprint{arXiv}{astro-ph/9712318}.

\bibitem{Bergstrom:2000pn}
Lars Bergstrom.
\newblock {Nonbaryonic dark matter: Observational evidence and detection
  methods}.
\newblock {\em Rept.Prog.Phys.}, 63:793, 2000,
  \doi{10.1088/0034-4885/63/5/2r3}, \eprint{arXiv}{hep-ph/0002126}.

\bibitem{Bringmann:2012vr}
Torsten Bringmann, Xiaoyuan Huang, Alejandro Ibarra, Stefan Vogl, and Christoph
  Weniger.
\newblock {Fermi LAT Search for Internal Bremsstrahlung Signatures from Dark
  Matter Annihilation}.
\newblock {\em JCAP}, 1207:054, 2012, \doi{10.1088/1475-7516/2012/07/054},
  \eprint{arXiv}{1203.1312}.

\bibitem{Weniger:2012tx}
Christoph Weniger.
\newblock {A Tentative Gamma-Ray Line from Dark Matter Annihilation at the
  Fermi Large Area Telescope}.
\newblock {\em JCAP}, 1208:007, 2012, \doi{10.1088/1475-7516/2012/08/007},
  \eprint{arXiv}{1204.2797}.

\bibitem{Bruch:2009rp}
Tobias Bruch, Annika~H.G. Peter, Justin Read, Laura Baudis, and George Lake.
\newblock {Dark Matter Disc Enhanced Neutrino Fluxes from the Sun and Earth}.
\newblock {\em Phys.Lett.}, B674:250\unskip--\ignorespaces 256, 2009,
  \doi{10.1016/j.physletb.2009.03.042}, \eprint{arXiv}{0902.4001}.

\bibitem{Hooper:2012sr}
Dan Hooper, Chris Kelso, and Farinaldo~S. Queiroz.
\newblock {Stringent and Robust Constraints on the Dark Matter Annihilation
  Cross Section From the Region of the Galactic Center}.
\newblock 2012, \eprint{arXiv}{1209.3015}.

\bibitem{Abazajian:2012pn}
Kevork~N. Abazajian and Manoj Kaplinghat.
\newblock {Detection of a Gamma-Ray Source in the Galactic Center Consistent
  with Extended Emission from Dark Matter Annihilation and Concentrated
  Astrophysical Emission}.
\newblock {\em Phys.Rev.}, D86:083511, 2012, \doi{10.1103/PhysRevD.86.083511},
  \eprint{arXiv}{1207.6047}.

\bibitem{Hooper:2013rwa}
Dan Hooper and Tracy~R. Slatyer.
\newblock {Two Emission Mechanisms in the Fermi Bubbles: A Possible Signal of
  Annihilating Dark Matter}.
\newblock 2013, \eprint{arXiv}{1302.6589}.

\bibitem{Hooper:2010im}
Dan Hooper and Tim Linden.
\newblock {Gamma Rays From The Galactic Center and the WMAP Haze}.
\newblock {\em Phys.Rev.}, D83:083517, 2011, \doi{10.1103/PhysRevD.83.083517},
  \eprint{arXiv}{1011.4520}.

\bibitem{Kaplinghat:2009ix}
Manoj Kaplinghat, Daniel~J. Phalen, and Kathryn~M. Zurek.
\newblock {Pulsars as the Source of the WMAP Haze}.
\newblock {\em JCAP}, 0912:010, 2009, \doi{10.1088/1475-7516/2009/12/010},
  \eprint{arXiv}{0905.0487}.

\bibitem{SiegalGaskins:2008ge}
Jennifer~M. Siegal-Gaskins.
\newblock {Revealing dark matter substructure with anisotropies in the diffuse
  gamma-ray background}.
\newblock {\em JCAP}, 0810:040, 2008, \doi{10.1088/1475-7516/2008/10/040},
  \eprint{arXiv}{0807.1328}.

\bibitem{Bai:2010hh}
Yang Bai, Patrick~J. Fox, and Roni Harnik.
\newblock {The Tevatron at the Frontier of Dark Matter Direct Detection}.
\newblock {\em JHEP}, 1012:048, 2010, \doi{10.1007/JHEP12(2010)048},
  \eprint{arXiv}{1005.3797}.

\bibitem{Carroll:2008ub}
Sean~M. Carroll, Sonny Mantry, Michael~J. Ramsey-Musolf, and Christoper~W.
  Stubbs.
\newblock {Dark-Matter-Induced Weak Equivalence Principle Violation}.
\newblock {\em Phys.Rev.Lett.}, 103:011301, 2009,
  \doi{10.1103/PhysRevLett.103.011301}, \eprint{arXiv}{0807.4363}.

\bibitem{Adelberger:2009zz}
E.G. Adelberger, J.H. Gundlach, B.R. Heckel, S.~Hoedl, and S.~Schlamminger.
\newblock {Torsion balance experiments: A low-energy frontier of particle
  physics}.
\newblock {\em Prog.Part.Nucl.Phys.}, 62:102\unskip--\ignorespaces 134, 2009,
  \doi{10.1016/j.ppnp.2008.08.002}.

\bibitem{Dobrescu:2006au}
Bogdan~A. Dobrescu and Irina Mocioiu.
\newblock {Spin-dependent macroscopic forces from new particle exchange}.
\newblock {\em JHEP}, 0611:005, 2006, \doi{10.1088/1126-6708/2006/11/005},
  \eprint{arXiv}{hep-ph/0605342}.

\bibitem{Hoedl:2011zz}
S.A. Hoedl, F.~Fleischer, E.G. Adelberger, and B.R. Heckel.
\newblock {Improved Constraints on an Axion-Mediated Force}.
\newblock {\em Phys.Rev.Lett.}, 106:041801, 2011,
  \doi{10.1103/PhysRevLett.106.041801}.

\bibitem{Cirigliano:2006dg}
Vincenzo Cirigliano, Stefano Profumo, and Michael~J. Ramsey-Musolf.
\newblock {Baryogenesis, Electric Dipole Moments and Dark Matter in the MSSM}.
\newblock {\em JHEP}, 0607:002, 2006, \doi{10.1088/1126-6708/2006/07/002},
  \eprint{arXiv}{hep-ph/0603246}.

\bibitem{McKeen:2013dma}
David McKeen, Maxim Pospelov, and Adam Ritz.
\newblock {EDM Signatures of PeV-scale Superpartners}.
\newblock 2013, \eprint{arXiv}{1303.1172}.

\bibitem{Feng_rev:2013}
Jonathan~L. Feng.
\newblock {Naturalness and the Status of Supersymmetry}.
\newblock 2013, \eprint{arXiv}{1302.6587}.

\bibitem{Jungman:1995df}
Gerard Jungman, Marc Kamionkowski, and Kim Griest.
\newblock {Supersymmetric dark matter}.
\newblock {\em Phys.Rept.}, 267:195\unskip--\ignorespaces 373, 1996,
  \doi{10.1016/0370-1573(95)00058-5}, \eprint{arXiv}{hep-ph/9506380}.

\bibitem{Servant:2002aq}
Geraldine Servant and Timothy~M.P. Tait.
\newblock {Is the lightest Kaluza-Klein particle a viable dark matter
  candidate?}
\newblock {\em Nucl.Phys.}, B650:391\unskip--\ignorespaces 419, 2003,
  \doi{10.1016/S0550-3213(02)01012-X}, \eprint{arXiv}{hep-ph/0206071}.

\bibitem{Cheng:2002ej}
Hsin-Chia Cheng, Jonathan~L. Feng, and Konstantin~T. Matchev.
\newblock {Kaluza-Klein dark matter}.
\newblock {\em Phys.Rev.Lett.}, 89:211301, 2002,
  \doi{10.1103/PhysRevLett.89.211301}, \eprint{arXiv}{hep-ph/0207125}.

\bibitem{Cembranos:2003mr}
J.A.R. Cembranos, A.~Dobado, and Antonio~Lopez Maroto.
\newblock {Brane world dark matter}.
\newblock {\em Phys.Rev.Lett.}, 90:241301, 2003,
  \doi{10.1103/PhysRevLett.90.241301}, \eprint{arXiv}{hep-ph/0302041}.

\bibitem{Goldberg:1983nd}
H.~Goldberg.
\newblock {Constraint on the Photino Mass from Cosmology}.
\newblock {\em Phys.Rev.Lett.}, 50:1419, 1983,
  \doi{10.1103/PhysRevLett.50.1419}.

\bibitem{Ellis:1983ew}
John~R. Ellis, J.S. Hagelin, Dimitri~V. Nanopoulos, Keith~A. Olive, and
  M.~Srednicki.
\newblock {Supersymmetric Relics from the Big Bang}.
\newblock {\em Nucl.Phys.}, B238:453\unskip--\ignorespaces 476, 1984,
  \doi{10.1016/0550-3213(84)90461-9}.

\bibitem{GeringerSameth:2011iw}
Alex Geringer-Sameth and Savvas~M. Koushiappas.
\newblock {Exclusion of canonical WIMPs by the joint analysis of Milky Way
  dwarfs with Fermi}.
\newblock {\em Phys.Rev.Lett.}, 107:241303, 2011,
  \doi{10.1103/PhysRevLett.107.241303}, \eprint{arXiv}{1108.2914}.

\bibitem{GeringerSameth:2012sr}
Alex Geringer-Sameth and Savvas~M. Koushiappas.
\newblock {Dark matter line search using a joint analysis of dwarf galaxies
  with the Fermi Gamma-ray Space Telescope}.
\newblock 2012, \eprint{arXiv}{1206.0796}.

\bibitem{Berger:2008cq}
Carola~F. Berger, James~S. Gainer, JoAnne~L. Hewett, and Thomas~G. Rizzo.
\newblock {Supersymmetry Without Prejudice}.
\newblock {\em JHEP}, 0902:023, 2009, \doi{10.1088/1126-6708/2009/02/023},
  \eprint{arXiv}{0812.0980}.

\bibitem{Dreiner:2007fw}
H.K. Dreiner, S.~Heinemeyer, O.~Kittel, U.~Langenfeld, A.M. Weber, et~al.
\newblock {How light can the lightest neutralino be?}
\newblock {\em eConf}, C0705302:SUS06, 2007, \eprint{arXiv}{0707.1425}.

\bibitem{Dreiner:2009ic}
Herbi~K. Dreiner, Sven Heinemeyer, Olaf Kittel, Ulrich Langenfeld, Arne~M.
  Weber, et~al.
\newblock {Mass Bounds on a Very Light Neutralino}.
\newblock {\em Eur.Phys.J.}, C62:547\unskip--\ignorespaces 572, 2009,
  \doi{10.1140/epjc/s10052-009-1042-y}, \eprint{arXiv}{0901.3485}.

\bibitem{Profumo:2008yg}
Stefano Profumo.
\newblock {Hunting the lightest lightest neutralinos}.
\newblock {\em Phys.Rev.}, D78:023507, 2008, \doi{10.1103/PhysRevD.78.023507},
  \eprint{arXiv}{0806.2150}.

\bibitem{Zurek:2008qg}
Kathryn~M. Zurek.
\newblock {Multi-Component Dark Matter}.
\newblock {\em Phys.Rev.}, D79:115002, 2009, \doi{10.1103/PhysRevD.79.115002},
  \eprint{arXiv}{0811.4429}.

\bibitem{Feng:2010ij}
Jonathan~L. Feng, Marc Kamionkowski, and Samuel~K. Lee.
\newblock {Light Gravitinos at Colliders and Implications for Cosmology}.
\newblock {\em Phys.Rev.}, D82:015012, 2010, \doi{10.1103/PhysRevD.82.015012},
  \eprint{arXiv}{1004.4213}.

\bibitem{Gardner:2010zf}
Susan Gardner.
\newblock {Implications of Standard-Model flavor violation for new physics
  searches}.
\newblock {\em AIP Conf.Proc.}, 1261:185\unskip--\ignorespaces 190, 2010,
  \doi{10.1063/1.3479341}, \eprint{arXiv}{1005.1366}.

\bibitem{Brodsky:2012zza}
Stanley Brodsky and Susan Gardner.
\newblock {The Impact of Intrinsic Heavy Quark Distributions in the Proton on
  New Physics Searches at the High Intensity Frontier}.
\newblock {\em SLAC-PUB-14828}, 2012.

\bibitem{Ellis:2008hf}
John~R. Ellis, Keith~A. Olive, and Christopher Savage.
\newblock {Hadronic Uncertainties in the Elastic Scattering of Supersymmetric
  Dark Matter}.
\newblock {\em Phys.Rev.}, D77:065026, 2008, \doi{10.1103/PhysRevD.77.065026},
  \eprint{arXiv}{0801.3656}.

\bibitem{Musolf:1993tb}
M.J. Musolf, T.W. Donnelly, J.~Dubach, S.J. Pollock, S.~Kowalski, et~al.
\newblock {Intermediate-energy semileptonic probes of the hadronic neutral
  current}.
\newblock {\em Phys.Rept.}, 239:1\unskip--\ignorespaces 178, 1994,
  \doi{10.1016/0370-1573(94)90040-X}.

\bibitem{Giedt:2009mr}
Joel Giedt, Anthony~W. Thomas, and Ross~D. Young.
\newblock {Dark matter, the CMSSM and lattice QCD}.
\newblock {\em Phys.Rev.Lett.}, 103:201802, 2009,
  \doi{10.1103/PhysRevLett.103.201802}, \eprint{arXiv}{0907.4177}.

\bibitem{Young:2013nn}
R.D. Young.
\newblock {Strange quark content of the nucleon and dark matter searches}.
\newblock {\em PoS}, LATTICE2012:014, 2012, \eprint{arXiv}{1301.1765}.

\bibitem{Junnarkar:2013ac}
Parikshit Junnarkar and Andre Walker-Loud.
\newblock {The Scalar Strange Content of the Nucleon from Lattice QCD}.
\newblock 2013, \eprint{arXiv}{1301.1114}.

\bibitem{Shifman:1978zn}
Mikhail~A. Shifman, A.I. Vainshtein, and Valentin~I. Zakharov.
\newblock {Remarks on Higgs Boson Interactions with Nucleons}.
\newblock {\em Phys.Lett.}, B78:443, 1978, \doi{10.1016/0370-2693(78)90481-1}.

\bibitem{Finkbeiner:2007kk}
Douglas~P. Finkbeiner and Neal Weiner.
\newblock {Exciting Dark Matter and the INTEGRAL/SPI 511 keV signal}.
\newblock {\em Phys.Rev.}, D76:083519, 2007, \doi{10.1103/PhysRevD.76.083519},
  \eprint{arXiv}{astro-ph/0702587}.

\bibitem{Finkbeiner:2009mi}
Douglas~P. Finkbeiner, Tracy~R. Slatyer, Neal Weiner, and Itay Yavin.
\newblock {PAMELA, DAMA, INTEGRAL and Signatures of Metastable Excited WIMPs}.
\newblock {\em JCAP}, 0909:037, 2009, \doi{10.1088/1475-7516/2009/09/037},
  \eprint{arXiv}{0903.1037}.

\bibitem{Hill:2011be}
Richard~J. Hill and Mikhail~P. Solon.
\newblock {Universal behavior in the scattering of heavy, weakly interacting
  dark matter on nuclear targets}.
\newblock {\em Phys.Lett.}, B707:539\unskip--\ignorespaces 545, 2012,
  \doi{10.1016/j.physletb.2012.01.013}, \eprint{arXiv}{1111.0016}.

\bibitem{Cirigliano:2012pq}
Vincenzo Cirigliano, Michael~L. Graesser, and Grigory Ovanesyan.
\newblock {WIMP-nucleus scattering in chiral effective theory}.
\newblock {\em JHEP}, 1210:025, 2012, \doi{10.1007/JHEP10(2012)025},
  \eprint{arXiv}{1205.2695}.

\bibitem{hoorabi}
Hooman Davoudiasl and Rabindra~N. Mohapatra.
\newblock {On Relating the Genesis of Cosmic Baryons and Dark Matter}.
\newblock {\em New J.Phys.}, 14:095011, 2012,
  \doi{10.1088/1367-2630/14/9/095011}, \eprint{arXiv}{1203.1247}.

\bibitem{Adriani:2008zq}
O.~Adriani, G.C. Barbarino, G.A. Bazilevskaya, R.~Bellotti, M.~Boezio, et~al.
\newblock {A new measurement of the antiproton-to-proton flux ratio up to 100
  GeV in the cosmic radiation}.
\newblock {\em Phys.Rev.Lett.}, 102:051101, 2009,
  \doi{10.1103/PhysRevLett.102.051101}, \eprint{arXiv}{0810.4994}.

\bibitem{Fox:2008kb}
Patrick~J. Fox and Erich Poppitz.
\newblock {Leptophilic Dark Matter}.
\newblock {\em Phys.Rev.}, D79:083528, 2009, \doi{10.1103/PhysRevD.79.083528},
  \eprint{arXiv}{0811.0399}.

\bibitem{Cholis:2008qq}
Ilias Cholis, Douglas~P. Finkbeiner, Lisa Goodenough, and Neal Weiner.
\newblock {The PAMELA Positron Excess from Annihilations into a Light Boson}.
\newblock {\em JCAP}, 0912:007, 2009, \doi{10.1088/1475-7516/2009/12/007},
  \eprint{arXiv}{0810.5344}.

\bibitem{ArkaniHamed:2008qn}
Nima Arkani-Hamed, Douglas~P. Finkbeiner, Tracy~R. Slatyer, and Neal Weiner.
\newblock {A Theory of Dark Matter}.
\newblock {\em Phys.Rev.}, D79:015014, 2009, \doi{10.1103/PhysRevD.79.015014},
  \eprint{arXiv}{0810.0713}.

\bibitem{Pospelov:2008jd}
Maxim Pospelov and Adam Ritz.
\newblock {Astrophysical Signatures of Secluded Dark Matter}.
\newblock {\em Phys.Lett.}, B671:391\unskip--\ignorespaces 397, 2009,
  \doi{10.1016/j.physletb.2008.12.012}, \eprint{arXiv}{0810.1502}.

\bibitem{Finkbeiner:2008qu}
Douglas~P. Finkbeiner, Tracy~R. Slatyer, and Neal Weiner.
\newblock {Nuclear scattering of dark matter coupled to a new light scalar}.
\newblock {\em Phys.Rev.}, D78:116006, 2008, \doi{10.1103/PhysRevD.78.116006},
  \eprint{arXiv}{0810.0722}.

\bibitem{Chao:2012mx}
Wei Chao, Matthew Gonderinger, and Michael~J. Ramsey-Musolf.
\newblock {Higgs Vacuum Stability, Neutrino Mass, and Dark Matter}.
\newblock {\em Phys.Rev.}, D86:113017, 2012, \doi{10.1103/PhysRevD.86.113017},
  \eprint{arXiv}{1210.0491}.

\bibitem{Bulava:2012rb}
John Bulava, Philipp Gerhold, Karl Jansen, Jim Kallarackal, Bastian
  Knippschild, et~al.
\newblock {Higgs-Yukawa model in chirally-invariant lattice field theory}.
\newblock 2012, \eprint{arXiv}{1210.1798}.

\bibitem{Patt:2006fw}
Brian Patt and Frank Wilczek.
\newblock {Higgs-field portal into hidden sectors}.
\newblock 2006, \eprint{arXiv}{hep-ph/0605188}.

\bibitem{Barger:2007im}
Vernon Barger, Paul Langacker, Mathew McCaskey, Michael~J. Ramsey-Musolf, and
  Gabe Shaughnessy.
\newblock {LHC Phenomenology of an Extended Standard Model with a Real Scalar
  Singlet}.
\newblock {\em Phys.Rev.}, D77:035005, 2008, \doi{10.1103/PhysRevD.77.035005},
  \eprint{arXiv}{0706.4311}.

\bibitem{FileviezPerez:2008bj}
Pavel Fileviez~Perez, Hiren~H. Patel, Michael.J. Ramsey-Musolf, and Kai Wang.
\newblock {Triplet Scalars and Dark Matter at the LHC}.
\newblock {\em Phys.Rev.}, D79:055024, 2009, \doi{10.1103/PhysRevD.79.055024},
  \eprint{arXiv}{0811.3957}.

\bibitem{Barger:2008jx}
Vernon Barger, Paul Langacker, Mathew McCaskey, Michael Ramsey-Musolf, and Gabe
  Shaughnessy.
\newblock {Complex Singlet Extension of the Standard Model}.
\newblock {\em Phys.Rev.}, D79:015018, 2009, \doi{10.1103/PhysRevD.79.015018},
  \eprint{arXiv}{0811.0393}.

\bibitem{Gonderinger:2009jp}
Matthew Gonderinger, Yingchuan Li, Hiren Patel, and Michael~J. Ramsey-Musolf.
\newblock {Vacuum Stability, Perturbativity, and Scalar Singlet Dark Matter}.
\newblock {\em JHEP}, 1001:053, 2010, \doi{10.1007/JHEP01(2010)053},
  \eprint{arXiv}{0910.3167}.

\bibitem{He:2009yd}
Xiao-Gang He, Tong Li, Xue-Qian Li, Jusak Tandean, and Ho-Chin Tsai.
\newblock {The Simplest Dark-Matter Model, CDMS II Results, and Higgs Detection
  at LHC}.
\newblock {\em Phys.Lett.}, B688:332\unskip--\ignorespaces 336, 2010,
  \doi{10.1016/j.physletb.2010.04.026}, \eprint{arXiv}{0912.4722}.

\bibitem{Cheung:2012nb}
Clifford Cheung, Michele Papucci, and Kathryn~M. Zurek.
\newblock {Higgs and Dark Matter Hints of an Oasis in the Desert}.
\newblock {\em JHEP}, 1207:105, 2012, \doi{10.1007/JHEP07(2012)105},
  \eprint{arXiv}{1203.5106}.

\bibitem{Gonderinger:2012rd}
Matthew Gonderinger, Hyungjun Lim, and Michael~J. Ramsey-Musolf.
\newblock {Complex Scalar Singlet Dark Matter: Vacuum Stability and
  Phenomenology}.
\newblock {\em Phys.Rev.}, D86:043511, 2012, \doi{10.1103/PhysRevD.86.043511},
  \eprint{arXiv}{1202.1316}.

\bibitem{Feng:2009mn}
Jonathan~L. Feng, Manoj Kaplinghat, Huitzu Tu, and Hai-Bo Yu.
\newblock {Hidden Charged Dark Matter}.
\newblock {\em JCAP}, 0907:004, 2009, \doi{10.1088/1475-7516/2009/07/004},
  \eprint{arXiv}{0905.3039}.

\bibitem{Ackerman:2008gi}
Lotty Ackerman, Matthew~R. Buckley, Sean~M. Carroll, and Marc Kamionkowski.
\newblock {Dark Matter and Dark Radiation}.
\newblock {\em Phys.Rev.}, D79:023519, 2009, \doi{10.1103/PhysRevD.79.023519},
  \eprint{arXiv}{0810.5126}.

\bibitem{Baumgart:2009tn}
Matthew Baumgart, Clifford Cheung, Joshua~T. Ruderman, Lian-Tao Wang, and Itay
  Yavin.
\newblock {Non-Abelian Dark Sectors and Their Collider Signatures}.
\newblock {\em JHEP}, 0904:014, 2009, \doi{10.1088/1126-6708/2009/04/014},
  \eprint{arXiv}{0901.0283}.

\bibitem{Holdom:1985ag}
Bob Holdom.
\newblock {Two U(1)'s and Epsilon Charge Shifts}.
\newblock {\em Phys.Lett.}, B166:196, 1986, \doi{10.1016/0370-2693(86)91377-8}.

\bibitem{Holdom:1986eq}
Bob Holdom.
\newblock {Searching for epsilon Charges and a New U(1)}.
\newblock {\em Phys.Lett.}, B178:65, 1986, \doi{10.1016/0370-2693(86)90470-3}.

\bibitem{Afanasev:2008fv}
A.~Afanasev, O.K. Baker, K.B. Beard, G.~Biallas, J.~Boyce, et~al.
\newblock {New Experimental Limit on Photon Hidden-Sector Paraphoton Mixing}.
\newblock {\em Phys.Lett.}, B679:317\unskip--\ignorespaces 320, 2009,
  \doi{10.1016/j.physletb.2009.07.055}, \eprint{arXiv}{0810.4189}.

\bibitem{Feldman:2007wj}
Daniel Feldman, Zuowei Liu, and Pran Nath.
\newblock {The Stueckelberg Z-prime Extension with Kinetic Mixing and
  Milli-Charged Dark Matter From the Hidden Sector}.
\newblock {\em Phys.Rev.}, D75:115001, 2007, \doi{10.1103/PhysRevD.75.115001},
  \eprint{arXiv}{hep-ph/0702123}.

\bibitem{Davidson:2000hf}
Sacha Davidson, Steen Hannestad, and Georg Raffelt.
\newblock {Updated bounds on millicharged particles}.
\newblock {\em JHEP}, 0005:003, 2000, \eprint{arXiv}{hep-ph/0001179}.

\bibitem{Gardner:2009et}
Susan Gardner and David~C. Latimer.
\newblock {Dark Matter Constraints from a Cosmic Index of Refraction}.
\newblock {\em Phys.Rev.}, D82:063506, 2010, \doi{10.1103/PhysRevD.82.063506},
  \eprint{arXiv}{0904.1612}.

\bibitem{Morganti:2011hg}
R.~Morganti et~al., LOFAR Collaboration.
\newblock {LOFAR: opening a new window on low frequency radio astronomy}.
\newblock 2011, \eprint{arXiv}{1112.5094}.

\bibitem{Ahlers:2007qf}
M.~Ahlers, H.~Gies, J.~Jaeckel, J.~Redondo, and A.~Ringwald.
\newblock {Laser experiments explore the hidden sector}.
\newblock {\em Phys.Rev.}, D77:095001, 2008, \doi{10.1103/PhysRevD.77.095001},
  \eprint{arXiv}{0711.4991}.

\bibitem{Masso:2006gc}
Eduard Masso and Javier Redondo.
\newblock {Compatibility of CAST search with axion-like interpretation of PVLAS
  results}.
\newblock {\em Phys.Rev.Lett.}, 97:151802, 2006,
  \doi{10.1103/PhysRevLett.97.151802}, \eprint{arXiv}{hep-ph/0606163}.

\bibitem{Kusenko:2001vu}
Alexander Kusenko and Paul~J. Steinhardt.
\newblock {Q ball candidates for selfinteracting dark matter}.
\newblock {\em Phys.Rev.Lett.}, 87:141301, 2001,
  \doi{10.1103/PhysRevLett.87.141301}, \eprint{arXiv}{astro-ph/0106008}.

\bibitem{Essig:2009nc}
Rouven Essig, Philip Schuster, and Natalia Toro.
\newblock {Probing Dark Forces and Light Hidden Sectors at Low-Energy e+e-
  Colliders}.
\newblock {\em Phys.Rev.}, D80:015003, 2009, \doi{10.1103/PhysRevD.80.015003},
  \eprint{arXiv}{0903.3941}.

\bibitem{Bjorken:2009mm}
James~D. Bjorken, Rouven Essig, Philip Schuster, and Natalia Toro.
\newblock {New Fixed-Target Experiments to Search for Dark Gauge Forces}.
\newblock {\em Phys.Rev.}, D80:075018, 2009, \doi{10.1103/PhysRevD.80.075018},
  \eprint{arXiv}{0906.0580}.

\bibitem{Fayet:2007ua}
Pierre Fayet.
\newblock {U-boson production in e+ e- annihilations, psi and Upsilon decays,
  and Light Dark Matter}.
\newblock {\em Phys.Rev.}, D75:115017, 2007, \doi{10.1103/PhysRevD.75.115017},
  \eprint{arXiv}{hep-ph/0702176}.

\bibitem{Pospelov:2008zw}
Maxim Pospelov.
\newblock {Secluded U(1) below the weak scale}.
\newblock {\em Phys.Rev.}, D80:095002, 2009, \doi{10.1103/PhysRevD.80.095002},
  \eprint{arXiv}{0811.1030}.

\bibitem{Okun:2006eb}
L.B. Okun.
\newblock {Mirror particles and mirror matter: 50 years of speculations and
  search}.
\newblock {\em Phys.Usp.}, 50:380\unskip--\ignorespaces 389, 2007,
  \doi{10.1070/PU2007v050n04ABEH006227}, \eprint{arXiv}{hep-ph/0606202}.

\bibitem{Berezhiani:2003xm}
Zurab Berezhiani.
\newblock {Mirror world and its cosmological consequences}.
\newblock {\em Int.J.Mod.Phys.}, A19:3775\unskip--\ignorespaces 3806, 2004,
  \doi{10.1142/S0217751X04020075}, \eprint{arXiv}{hep-ph/0312335}.

\bibitem{Nussinov:1985xr}
S.~Nussinov.
\newblock {Technocosmology: Could a Technibaryon Excess Provide a 'Natural'
  Missing Mass Candidate?}
\newblock {\em Phys.Lett.}, B165:55, 1985, \doi{10.1016/0370-2693(85)90689-6}.

\bibitem{Barr:1990ca}
Stephen~M. Barr, R.~Sekhar Chivukula, and Edward Farhi.
\newblock {Electroweak Fermion Number Violation and the Production of Stable
  Particles in the Early Universe}.
\newblock {\em Phys.Lett.}, B241:387\unskip--\ignorespaces 391, 1990,
  \doi{10.1016/0370-2693(90)91661-T}.

\bibitem{Kaplan:1991ah}
David~B. Kaplan.
\newblock {A Single explanation for both the baryon and dark matter densities}.
\newblock {\em Phys.Rev.Lett.}, 68:741\unskip--\ignorespaces 743, 1992,
  \doi{10.1103/PhysRevLett.68.741}.

\bibitem{zurek}
David~E. Kaplan, Markus~A. Luty, and Kathryn~M. Zurek.
\newblock {Asymmetric Dark Matter}.
\newblock {\em Phys.Rev.}, D79:115016, 2009, \doi{10.1103/PhysRevD.79.115016},
  \eprint{arXiv}{0901.4117}.

\bibitem{Gardner:2013aiw}
Susan Gardner and Daheng He.
\newblock {Radiative Beta Decay for Studies of CP Violation}.
\newblock 2013, \eprint{arXiv}{1302.1862}.

\bibitem{Gardner:2012rp}
Susan Gardner and Daheng He.
\newblock {A T-odd Momentum Correlation in Radiative $\beta$-Decay}.
\newblock {\em Phys.Rev.}, D86:016003, 2012, \doi{10.1103/PhysRevD.86.016003},
  \eprint{arXiv}{1202.5239}.

\bibitem{Bagnasco:1993st}
John Bagnasco, Michael Dine, and Scott~D. Thomas.
\newblock {Detecting technibaryon dark matter}.
\newblock {\em Phys.Lett.}, B320:99\unskip--\ignorespaces 104, 1994,
  \doi{10.1016/0370-2693(94)90830-3}, \eprint{arXiv}{hep-ph/9310290}.

\bibitem{Tulin:2012re}
Sean Tulin, Hai-Bo Yu, and Kathryn~M. Zurek.
\newblock {Oscillating Asymmetric Dark Matter}.
\newblock {\em JCAP}, 1205:013, 2012, \doi{10.1088/1475-7516/2012/05/013},
  \eprint{arXiv}{1202.0283}.

\bibitem{McKeown:2011yj}
R.D. McKeown.
\newblock {Electroweak Physics at Jefferson Lab}.
\newblock {\em AIP Conf.Proc.}, 1423:289\unskip--\ignorespaces 296, 2012,
  \doi{10.1063/1.3688816}, \eprint{arXiv}{1109.4855}.

\bibitem{Dudek:2012vr}
Jozef Dudek, Rolf Ent, Rouven Essig, K.S. Kumar, Curtis Meyer, et~al.
\newblock {Physics Opportunities with the 12 GeV Upgrade at Jefferson Lab}.
\newblock {\em Eur.Phys.J.}, A48:187, 2012, \doi{10.1140/epja/i2012-12187-1},
  \eprint{arXiv}{1208.1244}.

\bibitem{Aoyama:2012wj}
Tatsumi Aoyama, Masashi Hayakawa, Toichiro Kinoshita, and Makiko Nio.
\newblock {Tenth-Order QED Contribution to the Electron g-2 and an Improved
  Value of the Fine Structure Constant}.
\newblock {\em Phys.Rev.Lett.}, 109:111807, 2012,
  \doi{10.1103/PhysRevLett.109.111807}, \eprint{arXiv}{1205.5368}.

\bibitem{Hanneke:2008tm}
D.~Hanneke, S.~Fogwell, and G.~Gabrielse.
\newblock {New Measurement of the Electron Magnetic Moment and the Fine
  Structure Constant}.
\newblock {\em Phys.Rev.Lett.}, 100:120801, 2008,
  \doi{10.1103/PhysRevLett.100.120801}, \eprint{arXiv}{0801.1134}.

\bibitem{Bouchendira:2010es}
Rym Bouchendira, Pierre Clade, Saida Guellati-Khelifa, Francois Nez, and
  Francois Biraben.
\newblock {New determination of the fine structure constant and test of the
  quantum electrodynamics}.
\newblock {\em Phys.Rev.Lett.}, 106:080801, 2011,
  \doi{10.1103/PhysRevLett.106.080801}, \eprint{arXiv}{1012.3627}.

\bibitem{Davoudiasl:2012ig}
Hooman Davoudiasl, Hye-Sung Lee, and William~J. Marciano.
\newblock {Dark Side of Higgs Diphoton Decays and Muon g-2}.
\newblock {\em Phys.Rev.}, D86:095009, 2012, \doi{10.1103/PhysRevD.86.095009},
  \eprint{arXiv}{1208.2973}.

\bibitem{Minkowski:1977zr}
P.~{Minkowski}.
\newblock {{$\mu\to e\gamma$} at a rate of one out of 10$^{9}$ muon decays?}
\newblock {\em Physics Letters B}, 67:421\unskip--\ignorespaces 428, April
  1977, \doi{10.1016/0370-2693(77)90435-X}.

\bibitem{Yanagida:79}
T.~Yanagida.
\newblock {HORIZONTAL SYMMETRY AND MASSES OF NEUTRINOS}.
\newblock {\em in {\it Proceedings of the Workshop on the Unified Theory and
  the Baryon Number in the Universe}, (O. Sawada and A. Sugamoto, eds.), KEK,
  Tsukuba, Japan}, C7902131(Tsukuba, Japan):95, 1979.

\bibitem{Gell-Mann:80}
M.~{Gell-Mann}, P.~{Ramond}, and R.~{Slansky}.
\newblock {COMPLEX SPINORS AND UNIFIED THEORIES}.
\newblock {\em in {\it Supergravity}, (P. van Nieuwenhuizen et al. eds.), North
  Holland, Amsterdam}, 1980.

\bibitem{Glashow:1980}
S.~{Glashow}.
\newblock see-saw.
\newblock {\em in {\it Proceedings of the 1979 Cargese Summer Institute on
  Quarks and Leptons}, (M. Levy et al. eds.), Plenum Press, New York}, 1980.

\bibitem{Mohapatra:1980mz}
R.~N. {Mohapatra} and G.~{Senjanovic}.
\newblock {Neutrino mass and spontaneous parity nonconservation}.
\newblock {\em Physical Review Letters}, 44:912\unskip--\ignorespaces 915,
  April 1980, \doi{10.1103/PhysRevLett.44.912}.

\bibitem{Kusenko:2009lr}
A.~{Kusenko}.
\newblock {Sterile neutrinos: The dark side of the light fermions}.
\newblock {\em Physics Reports}, 481:1\unskip--\ignorespaces 28, September
  2009, \doi{10.1016/j.physrep.2009.07.004}, \eprint{arXiv}{0906.2968}.

\bibitem{de-Gouvea:2005fk}
A.~{de Gouv{\^e}a}.
\newblock {Seesaw energy scale and the LSND anomaly}.
\newblock {\em Phys.Rev. D}, 72(3):033005\unskip--\ignorespaces +, August 2005,
  \doi{10.1103/PhysRevD.72.033005}, \eprint{}{arXiv:hep-ph/0501039}.

\bibitem{de-Gouvea:2007lr}
A.~{de Gouv{\^e}a}, J.~{Jenkins}, and N.~{Vasudevan}.
\newblock {Neutrino phenomenology of very low-energy seesaw scenarios}.
\newblock {\em Phys.Rev. D}, 75(1):013003\unskip--\ignorespaces +, January
  2007, \doi{10.1103/PhysRevD.75.013003}, \eprint{}{arXiv:hep-ph/0608147}.

\bibitem{Kusenko:2010qy}
A.~{Kusenko}, F.~{Takahashi}, and T.~T. {Yanagida}.
\newblock {Dark matter from split seesaw}.
\newblock {\em Physics Letters B}, 693:144\unskip--\ignorespaces 148, September
  2010, \doi{10.1016/j.physletb.2010.08.031}, \eprint{arXiv}{1006.1731}.

\bibitem{Dodelson:1994rt}
S.~{Dodelson} and L.~M. {Widrow}.
\newblock {Sterile neutrinos as dark matter}.
\newblock {\em Physical Review Letters}, 72:17\unskip--\ignorespaces 20,
  January 1994, \doi{10.1103/PhysRevLett.72.17},
  \eprint{}{arXiv:hep-ph/9303287}.

\bibitem{Shi:1999lq}
X.~{Shi} and G.~M. {Fuller}.
\newblock {New Dark Matter Candidate: Nonthermal Sterile Neutrinos}.
\newblock {\em Physical Review Letters}, 82:2832\unskip--\ignorespaces 2835,
  April 1999, \doi{10.1103/PhysRevLett.82.2832},
  \eprint{}{arXiv:astro-ph/9810076}.

\bibitem{Abazajian:2001lr}
K.~{Abazajian}, G.~M. {Fuller}, and M.~{Patel}.
\newblock {Sterile neutrino hot, warm, and cold dark matter}.
\newblock {\em Phys.Rev. D}, 64(2):023501\unskip--\ignorespaces +, July 2001,
  \doi{10.1103/PhysRevD.64.023501}, \eprint{}{arXiv:astro-ph/0101524}.

\bibitem{Dolgov:2002ve}
A.~D. {Dolgov} and S.~H. {Hansen}.
\newblock {Massive sterile neutrinos as warm dark matter}.
\newblock {\em Astroparticle Physics}, 16:339\unskip--\ignorespaces 344,
  January 2002, \doi{10.1016/S0927-6505(01)00115-3},
  \eprint{}{arXiv:hep-ph/0009083}.

\bibitem{Abazajian:2002bh}
K.~N. {Abazajian} and G.~M. {Fuller}.
\newblock {Bulk QCD thermodynamics and sterile neutrino dark matter}.
\newblock {\em Phys.Rev. D}, 66(2):023526\unskip--\ignorespaces +, July 2002,
  \doi{10.1103/PhysRevD.66.023526}, \eprint{}{arXiv:astro-ph/0204293}.

\bibitem{Asaka:2005fj}
T.~{Asaka}, S.~{Blanchet}, and M.~{Shaposhnikov}.
\newblock {The {$\nu$}MSM, dark matter and neutrino masses [rapid
  communication]}.
\newblock {\em Physics Letters B}, 631:151\unskip--\ignorespaces 156, December
  2005, \doi{10.1016/j.physletb.2005.09.070}, \eprint{}{arXiv:hep-ph/0503065}.

\bibitem{Abazajian:2006qf}
K.~{Abazajian}.
\newblock {Production and evolution of perturbations of sterile neutrino dark
  matter}.
\newblock {\em Phys.Rev. D}, 73(6):063506\unskip--\ignorespaces +, March 2006,
  \doi{10.1103/PhysRevD.73.063506}, \eprint{}{arXiv:astro-ph/0511630}.

\bibitem{Shaposhnikov:2006fj}
M.~{Shaposhnikov} and I.~{Tkachev}.
\newblock {The {$\nu$}MSM, inflation, and dark matter}.
\newblock {\em Physics Letters B}, 639:414\unskip--\ignorespaces 417, August
  2006, \doi{10.1016/j.physletb.2006.06.063}, \eprint{}{arXiv:hep-ph/0604236}.

\bibitem{Boyanovsky:2007lr}
D.~{Boyanovsky} and C.-M. {Ho}.
\newblock {Sterile neutrino production via active-sterile oscillations: the
  quantum Zeno effect}.
\newblock {\em Journal of High Energy Physics}, 7:30\unskip--\ignorespaces +,
  July 2007, \doi{10.1088/1126-6708/2007/07/030},
  \eprint{}{arXiv:hep-ph/0612092}.

\bibitem{Boyanovsky:2007fk}
D.~{Boyanovsky}.
\newblock {Production of a sterile species via active-sterile mixing: An
  exactly solvable model}.
\newblock {\em Phys.Rev. D}, 76(10):103514\unskip--\ignorespaces +, November
  2007, \doi{10.1103/PhysRevD.76.103514}, \eprint{arXiv}{0706.3167}.

\bibitem{Shaposhnikov:2007qy}
M.~{Shaposhnikov}.
\newblock {A possible symmetry of the {$\nu$}MSM}.
\newblock {\em Nuclear Physics B}, 763:49\unskip--\ignorespaces 59, February
  2007, \doi{10.1016/j.nuclphysb.2006.11.003}, \eprint{}{arXiv:hep-ph/0605047}.

\bibitem{Gorbunov:2007uq}
D.~{Gorbunov} and M.~{Shaposhnikov}.
\newblock {How to find neutral leptons of the {$\nu$}MSM?}
\newblock {\em Journal of High Energy Physics}, 10:15\unskip--\ignorespaces +,
  October 2007, \doi{10.1088/1126-6708/2007/10/015}, \eprint{arXiv}{0705.1729}.

\bibitem{Kishimoto:2008pd}
C.~T. {Kishimoto} and G.~M. {Fuller}.
\newblock {Lepton-number-driven sterile neutrino production in the early
  universe}.
\newblock {\em Phys.Rev. D}, 78(2):023524\unskip--\ignorespaces +, July 2008,
  \doi{10.1103/PhysRevD.78.023524}, \eprint{arXiv}{0802.3377}.

\bibitem{Laine:2008kx}
M.~{Laine} and M.~{Shaposhnikov}.
\newblock {Sterile neutrino dark matter as a consequence of {$\nu$}MSM-induced
  lepton asymmetry}.
\newblock {\em JCAP}, 6:31\unskip--\ignorespaces +, June 2008,
  \doi{10.1088/1475-7516/2008/06/031}, \eprint{arXiv}{0804.4543}.

\bibitem{Petraki:2008yq}
K.~{Petraki}.
\newblock {Small-scale structure formation properties of chilled sterile
  neutrinos as dark matter}.
\newblock {\em Phys.Rev. D}, 77(10):105004\unskip--\ignorespaces +, May 2008,
  \doi{10.1103/PhysRevD.77.105004}, \eprint{arXiv}{0801.3470}.

\bibitem{Petraki:2008vn}
K.~{Petraki} and A.~{Kusenko}.
\newblock {Dark-matter sterile neutrinos in models with a gauge singlet in the
  Higgs sector}.
\newblock {\em Phys.Rev. D}, 77(6):065014\unskip--\ignorespaces +, March 2008,
  \doi{10.1103/PhysRevD.77.065014}, \eprint{arXiv}{0711.4646}.

\bibitem{Biermann:2006mz}
P.~L. {Biermann} and A.~{Kusenko}.
\newblock {Relic keV Sterile Neutrinos and Reionization}.
\newblock {\em Physical Review Letters}, 96(9):091301\unskip--\ignorespaces +,
  March 2006, \doi{10.1103/PhysRevLett.96.091301},
  \eprint{}{arXiv:astro-ph/0601004}.

\bibitem{Mapelli:2006gf}
M.~{Mapelli}, A.~{Ferrara}, and E.~{Pierpaoli}.
\newblock {Impact of dark matter decays and annihilations on reionization}.
\newblock {\em Mon. Not. Royal Astron. Soc.}, 369:1719\unskip--\ignorespaces
  1724, July 2006, \doi{10.1111/j.1365-2966.2006.10408.x},
  \eprint{}{arXiv:astro-ph/0603237}.

\bibitem{Stasielak:2007ly}
J.~{Stasielak}, P.~L. {Biermann}, and A.~{Kusenko}.
\newblock {Thermal Evolution of the Primordial Clouds in Warm Dark Matter
  Models with keV Sterile Neutrinos}.
\newblock {\em Astrophys.J.}, 654:290\unskip--\ignorespaces 303, January 2007,
  \doi{10.1086/509066}, \eprint{}{arXiv:astro-ph/0606435}.

\bibitem{Kusenko:1999rt}
A.~{Kusenko} and G.~{Segr{\`e}}.
\newblock {Pulsar kicks from neutrino oscillations}.
\newblock {\em Phys.Rev. D}, 59(6):061302\unskip--\ignorespaces +, March 1999,
  \doi{10.1103/PhysRevD.59.061302}, \eprint{}{arXiv:astro-ph/9811144}.

\bibitem{Barkovich:2004ul}
M.~{Barkovich}, J.~C. {D'Olivo}, and R.~{Montemayor}.
\newblock {Active-sterile neutrino oscillations and pulsar kicks}.
\newblock {\em Phys.Rev. D}, 70(4):043005\unskip--\ignorespaces +, August 2004,
  \doi{10.1103/PhysRevD.70.043005}, \eprint{}{arXiv:hep-ph/0402259}.

\bibitem{Fuller:2003vn}
G.~M. {Fuller}, A.~{Kusenko}, I.~{Mocioiu}, and S.~{Pascoli}.
\newblock {Pulsar kicks from a dark-matter sterile neutrino}.
\newblock {\em Phys.Rev. D}, 68(10):103002\unskip--\ignorespaces +, November
  2003, \doi{10.1103/PhysRevD.68.103002}, \eprint{}{arXiv:astro-ph/0307267}.

\bibitem{Loveridge:2004fr}
L.~C. {Loveridge}.
\newblock {Effects of Gravity and Finite Temperature on the Decay of the False
  Vacuum}.
\newblock {\em ArXiv High Energy Physics - Theory e-prints}, September 2004,
  \eprint{}{arXiv:hep-th/0409093}.

\bibitem{Kishimoto:2011fk}
C.~T. {Kishimoto}.
\newblock {Pulsar Kicks from Active-Sterile Neutrino Transformation in
  Supernovae}.
\newblock {\em ArXiv e-prints}, January 2011, \eprint{arXiv}{1101.1304}.

\bibitem{Akhmedov:1998ys}
E.~K. {Akhmedov}, V.~A. {Rubakov}, and A.~Y. {Smirnov}.
\newblock {Baryogenesis via Neutrino Oscillations}.
\newblock {\em Physical Review Letters}, 81:1359\unskip--\ignorespaces 1362,
  August 1998, \doi{10.1103/PhysRevLett.81.1359},
  \eprint{}{arXiv:hep-ph/9803255}.

\bibitem{Asaka:2005fr}
T.~{Asaka} and M.~{Shaposhnikov}.
\newblock {The @nMSM, dark matter and baryon asymmetry of the universe [rapid
  communication]}.
\newblock {\em Physics Letters B}, 620:17\unskip--\ignorespaces 26, July 2005,
  \doi{10.1016/j.physletb.2005.06.020}, \eprint{}{arXiv:hep-ph/0505013}.

\bibitem{Hidaka:2006yq}
J.~{Hidaka} and G.~M. {Fuller}.
\newblock {Dark matter sterile neutrinos in stellar collapse: Alteration of
  energy/lepton number transport, and a mechanism for supernova explosion
  enhancement}.
\newblock {\em Phys.Rev. D}, 74(12):125015\unskip--\ignorespaces +, December
  2006, \doi{10.1103/PhysRevD.74.125015}, \eprint{}{arXiv:astro-ph/0609425}.

\bibitem{Fryer:2006rt}
C.~L. {Fryer} and A.~{Kusenko}.
\newblock {Effects of Neutrino-driven Kicks on the Supernova Explosion
  Mechanism}.
\newblock {\em Astrophs. J. Suppl.}, 163:335\unskip--\ignorespaces 343, April
  2006, \doi{10.1086/500933}, \eprint{}{arXiv:astro-ph/0512033}.

\bibitem{Hidaka:2007kx}
J.~{Hidaka} and G.~M. {Fuller}.
\newblock {Sterile neutrino-enhanced supernova explosions}.
\newblock {\em Phys.Rev. D}, 76(8):083516\unskip--\ignorespaces +, October
  2007, \doi{10.1103/PhysRevD.76.083516}, \eprint{arXiv}{0706.3886}.

\bibitem{Fuller:2009uq}
G.~M. {Fuller}, A.~{Kusenko}, and K.~{Petraki}.
\newblock {Heavy sterile neutrinos and supernova explosions}.
\newblock {\em Physics Letters B}, 670:281\unskip--\ignorespaces 284, January
  2009, \doi{10.1016/j.physletb.2008.11.016}, \eprint{arXiv}{0806.4273}.

\bibitem{Joudaki:2012uk}
Shahab Joudaki, Kevork~N. Abazajian, and Manoj Kaplinghat.
\newblock {Are Light Sterile Neutrinos Preferred or Disfavored by Cosmology?}
\newblock {\em Phys. Rev.}, D87:065003, 2013, \eprint{arXiv}{1208.4354}.

\bibitem{Abazajian:2001fk}
K.~{Abazajian}, G.~M. {Fuller}, and W.~H. {Tucker}.
\newblock {Direct Detection of Warm Dark Matter in the X-Ray}.
\newblock {\em Astrophys.J.}, 562:593\unskip--\ignorespaces 604, December 2001,
  \doi{10.1086/323867}, \eprint{}{arXiv:astro-ph/0106002}.

\bibitem{Yuksel:2008fk}
H.~{Y{\"u}ksel}, J.~F. {Beacom}, and C.~R. {Watson}.
\newblock {Strong Upper Limits on Sterile Neutrino Warm Dark Matter}.
\newblock {\em Physical Review Letters}, 101(12):121301, September 2008,
  \doi{10.1103/PhysRevLett.101.121301}, \eprint{arXiv}{0706.4084}.

\bibitem{McLaughlin:1999fk}
G.~C. {McLaughlin}, J.~M. {Fetter}, A.~B. {Balantekin}, and G.~M. {Fuller}.
\newblock {Active-sterile neutrino transformation solution for r-process
  nucleosynthesis}.
\newblock {\em Phys.Rev. C}, 59:2873\unskip--\ignorespaces 2887, May 1999,
  \doi{10.1103/PhysRevC.59.2873}, \eprint{}{arXiv:astro-ph/9902106}.

\bibitem{Caldwell:2000db}
D.~O. {Caldwell}, G.~M. {Fuller}, and {Y.-Z.} {Qian}.
\newblock {Sterile neutrinos and supernova nucleosynthesis}.
\newblock {\em Phys.Rev. D}, 61(12):123005\unskip--\ignorespaces +, June 2000,
  \doi{10.1103/PhysRevD.61.123005}, \eprint{}{arXiv:astro-ph/9910175}.

\bibitem{Fetter:2003lr}
J.~{Fetter}, G.~C. {McLaughlin}, A.~B. {Balantekin}, and G.~M. {Fuller}.
\newblock {Active-sterile neutrino conversion: consequences for the r-process
  and supernova neutrino detection}.
\newblock {\em Astroparticle Physics}, 18:433\unskip--\ignorespaces 448,
  February 2003, \doi{10.1016/S0927-6505(02)00156-1},
  \eprint{}{arXiv:hep-ph/0205029}.

\end{thebibliography}

\end{document}